\documentclass[lineno]{jfm_older}
\usepackage{graphicx}
\usepackage{newtxtext}
\usepackage{newtxmath}
\usepackage{bm}
\usepackage{siunitx}
\usepackage{amsfonts}
\usepackage{graphicx}
\usepackage{amsmath}
\usepackage{mathtools}  
\usepackage{diffcoeff}  
\usepackage{xcolor}
\usepackage{natbib}
\usepackage{hyperref}
\hypersetup{
    colorlinks = true,
    urlcolor   = blue,
    citecolor  = black,
}

\title{Surface reflection of bottom generated oceanic lee waves}

\author{L. E. Baker\aff{1}
\corresp{\email{l.baker18@imperial.ac.uk}},
\and A. Mashayek \aff{1}}

\affiliation{\aff{1}Department of Civil and Environmental Engineering, Imperial College London}

\begin{document}
\maketitle

\begin{abstract}
Lee waves generated by stratified flow over rough bottom topography in the ocean extract momentum and energy from the geostrophic flow, causing drag and enhancing turbulence and mixing in the interior ocean when they break. Inviscid linear theory is generally used to predict the generation rate of lee waves, but the location and mechanism of wave breaking leading to eventual dissipation of energy and irreversible mixing are poorly constrained. In this study, a linear model with viscosity, diffusivity, and an upper boundary is used to demonstrate the potential importance of the surface in reflecting lee wave energy back into the interior, making the case for treating lee waves as a full water column process. In the absence of critical levels, it is shown that lee waves can be expected to interact with the upper ocean, resulting in enhanced vertical velocities and dissipation and mixing near the surface. The impact of the typical oceanic conditions of increasing background velocity and stratification with height above bottom are investigated and shown to contribute to enhanced upper ocean vertical velocities and mixing.
\end{abstract}

%

\section{Introduction} \label{sec:intro}
Oceanic lee waves are quasi-steady internal gravity waves generated by the interaction of geostrophic flows with submarine topography. They are present throughout the world's oceans, accounting for an estimated $0.2$ - $\SI{0.75}{\tera\watt}$ of conversion from the mean flow \citep{Scott2011,Nikurashin2011a,Wright2014}. Approximately half of this generation takes place in the Southern Ocean (SO) \citep{Nikurashin2011a}, where lee waves have been shown to be an important sink of energy and momentum from the energetic mesoscale eddies of the Antarctic Circumpolar Current (ACC) due to the rough topography and high bottom velocities in the region  \citep{Nikurashin2010b,Nikurashin2012a, NaveiraGarabato2004a, Yang2018a}. 

Lee waves play an important role not only in the momentum budget of the mean flow through lee wave drag, but also in the buoyancy and tracer budgets through diapycnal mixing. Enhanced levels of turbulence above topography associated with lee waves and other topographic interaction processes are an important source of diapycnal mixing in the deep ocean, contributing to the closure of the meridional overturning circulation (MOC) \citep{Mackinnon2017,Cessi2019, Cimoli2021}. The Southern Ocean upwelling of tracers such as $\mathrm{CO_2}$ and nutrients for primary production are also sensitive to mixing in the ocean, with important consequences for air-sea fluxes and ultimately climate \citep{Talley2016}. 

Lee wave horizontal lengthscales are typically of order $\SI{500}{\metre}$  - $\SI{10}{\kilo\metre}$ in the ocean, a range that is unresolved in global climate models, so the mixing and drag effects of lee waves both need to be parametrised. The generation of lee waves is usually understood using linear theory, whereby the lee wave perturbations are assumed to have a much smaller amplitude than the mean flow itself \citep{Bell1975}. An important parameter determining the linearity of lee waves generated at topography of characteristic height $h$ in uniform background stratification $N$ and velocity $U$ is the lee wave Froude number $Fr_L = Nh/U$  \citep{Mayer2017}. Lee waves can propagate vertically when their horizontal wavenumber $k$ (set by topography) is such that $|f| < |Uk| < |N|$, where $f$ is the Coriolis parameter. Under the assumption $|f| \ll |Uk| \ll |N|$, $Fr_L$ is proportional to the ratio of the topographic height $h$ to the lee wave vertical wavelength, or equivalently the ratio of the perturbation horizontal velocity to the background velocity, thus the linear approximation is formally valid for $Fr_L \ll 1$. Energy flux calculated using the linear approximation has been shown to agree with two-dimensional (2D) nonlinear simulations for $Fr_L \lesssim \mathcal{O}(1)$ \citep{Nikurashin2014}. 

For 2D topography and flow conditions such that $Fr_L$ is greater than some critical Froude number $Fr_L^{\text{crit}} \sim \mathcal{O}(1)$, topographic blocking occurs since the flow lacks the kinetic energy to raise itself over a bump of height greater than $\sim U/N$ \citep{Smith1989}. Thus, the effective height of topography $h^{\text{eff}}$ is always reduced such that the waves are generated with $Fr_L^{\text{eff}} = Nh^{\text{eff}}/U \lesssim Fr_L^{\text{crit}}$ \citep{Winters2012}. When the topography is three dimensional (3D), splitting may also occur as the flow goes around rather than over a bump, and the effective height is lower still. \citet{Nikurashin2014} found that for multichromatic topography with $h$ defined as the RMS (root mean square) topographic height, $Fr_L^{\text{crit}} \simeq 0.7$ for 2D topography, and $Fr_L^{\text{crit}} \simeq 0.4$ for 3D topography. Thus, with modifications for finite amplitude and 3D effects, the linear theory can be used with some success even when the topography is nonlinear. Several estimates of energy conversion from the geostrophic flow to lee waves have been found using estimated topographic spectra, bottom velocities and stratification globally \citep{Bell1975, Scott2011,Nikurashin2011a,Wright2014}. Problems remain with this approach, such as the proper representation of blocking in the topographic spectrum, and the neglect of the influence of flow due to large scale topography on the radiating lee waves, which can significantly impact the dissipation above topography \citep{Klymak2018}.

Although the generation of lee waves is well understood in a linear sense, the ultimate fate of lee wave energy as a fundamentally nonlinear and dissipative process is poorly constrained. After generation, lee waves radiate vertically and downstream away from topography. A vertical structure function exponentially decreasing with height above bottom was proposed by \cite{St.Laurent2002a} for parametrisation of dissipation rate due to the internal tide, and this has been implemented in lee wave parametrisations with decay scales between $300$ and $\SI{1000}{\metre}$ \citep{Nikurashin2013,Melet2014}. Both of these studies found that water mass transformation was sensitive to the decay scale used, thus accurate parametrisation of the vertical structure of mixing and dissipation is necessary for correctly predicting the ocean state in global climate models. 

Lee waves also play an important role in causing drag on the mean flow. When flow impinges on topography, the pressure difference across the topographic features cause drag, known as form drag. If there is topographic blocking, or the conditions for radiation of lee waves are not met, this drag will force the flow local to the topography. However, if lee waves are generated and propagate upwards this drag is distributed across the water column as a wave drag, locally forcing the flow where the waves break. Thus, the vertical distribution of the decelerating force on the mean flow due to lee wave breaking must also be parametrised. In the linear theory, the total lee wave drag is equal to the energy flux at topography multiplied by the bottom background velocity, but the vertical distributions of the forcing on the flow and the energy loss need not be the same.

Possible sinks for lee wave energy include breaking due to vertical shear from inertial oscillations generated by parametric instability \citep{Nikurashin2010a}, dissipation at critical levels \citep{Booker1967}, breaking due to convective instability on generation \citep{Peltier1979}, and re-absorption of lee wave energy in a shear flow \citep{Kunze2019}. \citet{Nikurashin2010b} performed idealised simulations representative of lee wave generation and dissipation in the Southern Ocean, finding that 50\% of lee wave energy dissipated in the bottom 1km of the ocean for $Fr_L \geq 0.5$ compared to 10\%  for $Fr_L = 0.2$. A more realistic simulation capturing the characteristic stratification, wind forcing, and topography of the SO \citep{Nikurashin2012a} found that 80\% of the wind power input into geostrophic eddies was converted to smaller scales by topography, of which just 20\% radiated into the interior ocean, with most dissipated in the bottom $\SI{100}{\metre}$. However, this and other wave resolving models may use artifically high diffusivity and viscosity, preventing lee waves from radiating in a physical way \citep{Shakespeare2017a}. 

The linear theory of \citet{Bell1975} uses a freely radiating upper boundary condition (hereafter referred to as `unbounded' theory), and can only be applied for uniform stratification and velocity, or by using the WKBJ approximation \citep{Gill1982}. This has led most idealized ocean lee wave studies to assume the same and treat lee waves as a process confined to the deep ocean where stratification and velocity are assumed to be approximately constant with height. The assumption in most such studies (with some exceptions, e.g. \citet{Zheng2019}) is that no significant amount of lee wave energy reaches the surface, and even if it does, it does not matter for the structure of the wave field. In this study, we consider the treatment of lee waves as a full water column process, allowing reflection from the surface and interaction with changes in stratification and velocity with height. 

In the real, dissipative ocean, some lee wave energy will be lost immediately due to boundary processes, and on their passage through the water column lee waves can be expected to lose energy through nonlinear processes leading to cascade of energy to smaller and eventually dissipative scales. Any model that tries to capture the entire water column must therefore include some representation of mixing and dissipation. However, the question of the magnitude and location of lee wave energy loss is a circular one, since it is the nonlinear interactions involving the wave field itself that cause wave breaking, leading to mixing and dissipation. Parametrisations for energy loss must therefore be used even when the lee waves are resolved, since capturing the lengthscales of both lee waves ($\sim \mathcal{O}(\SI{5}{\kilo\metre})$) and turbulent lengthscales ($\sim \mathcal{O}(\SI{1}{\centi\metre})$) in a 3D direct numerical simulation (DNS) remains prohibitively expensive. \citet{Shakespeare2017a} investigated the impact of Laplacian parametrisation of mixing and dissipation in lee wave resolving models, and concluded that care must be taken to avoid artificially high viscosity and diffusivity that is not physically justified. They suggest that high levels of dissipation near the bottom boundary in wave resolving models could be a direct result of the high levels of viscosity and diffusivity used in the sub-gridscale parametrisation. Therefore, lee wave dissipation in the abyssal ocean could be commonly overestimated in modelling studies, preventing the radiation of lee wave energy far up into the water column.

Observations of lee waves are sparse due to their unpredictable generation by the time varying eddy field, difficulty in taking measurements at the bottom of the ocean, and their steady nature \citep{Legg2020}. However, the available observational evidence indicates that linear predictions of energy flux overestimate the levels of dissipation in the bottom $\SI{1}{\kilo\metre}$ by up to an order of magnitude \citep{Brearley2013,Sheen2013,Waterman2013}. Direct measurements of lee wave energy flux over the Shackleton fracture zone in the Drake Passage \citep{Cusack2017} were found to be consistent with predicted linear generation modified for finite amplitude topography, but dissipation integrated over the water column was found to be two orders of magnitude smaller than expected, suggesting that lee waves find a sink for their energy outside of local mixing and dissipation.

One possible sink is reabsorption of lee wave energy to a sheared mean flow when the flow is decreasing in magnitude away from topography \citep{Kunze2019}. This is particularly relevant in regions of enhanced bottom velocities, and is supported by observational evidence that locations of overpredicted lee wave dissipation rates in the ACC are characterised by large near-bottom velocities \citep{Waterman2014}. \citet{Zheng2019} investigated another possible pathway, showing that that lee wave energy can be swept downstream to dissipate elsewhere. An important component to their study is an upper boundary, which allows lee waves at scales affected by rotation or nonhydrostatic effects to travel downstream by first reflecting at the upper boundary. They find that wave reflection enhances energy dissipation rates in the interior by up to a factor of two. 
\begin{figure} 
\centering
\includegraphics[width=\textwidth]{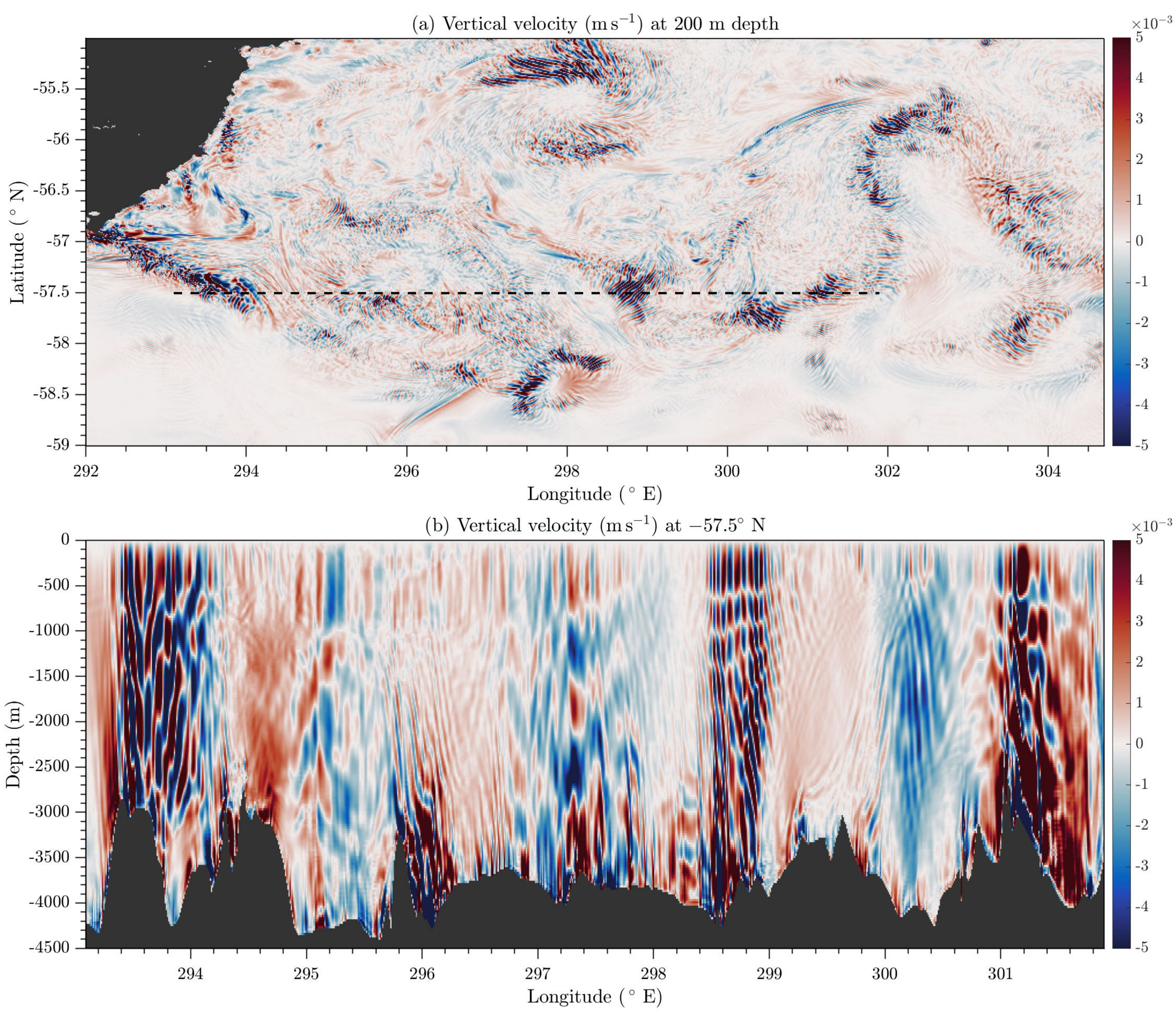} \\
\caption{A daily average of vertical velocity (\SI{}{\metre\per\second}) in a realistic simulation of the Drake Passage showing a strong lee wave field throughout the water column (details in main text). (a) A plan view at $\SI{200}{\metre}$, and (b) a vertical slice through the dashed line in (a).}\label{fig1} 
\end{figure}

The motivation for the current study arises from realistic regional simulations of the Southern Ocean that show large lee waves penetrating high into the water column and reflecting from the surface. Figure \ref{fig1} shows vertical velocities from a recent nested simulation of the Drake Passage at $0.01^\circ$ resolution, performed using the hydrostatic configuration of the Massachusetts Institute of Technology general circulation model (MITgcm, \citealt{Marshall1997}). For details of the model setup see \citet{Mashayek2017a} - the model shown here has an improvement of vertical resolution from $100$ to $225$ vertical levels, with $\SI{10}{\metre}$ resolution at the surface and $\leq \SI{25}{\metre}$ for all depths above $\SI{-4500}{\metre}$, allowing better resolution of the energetic internal wave field. The vertical diffusivity and viscosity have background values of $\SI{5e-5}{\metre\squared\per\second}$, and are enhanced by the $K$-profile parametrisation with the critical Richardson number for shear instability set to $Ri_c = 0.3$  \citep{Large1994}. Biharmonic Leith horizontal viscosity is used with a coefficient of $2$ \citep{Leith1996,Fox-Kemper2008}.

Figure \ref{fig1}a shows a plan view of a typical daily average of vertical velocity at $\SI{200}{\metre}$ depth. Lee waves appear as disturbances in the vertical velocity with $\mathcal{O}(\SI{0.1}{\degree}) \sim \mathcal{O}(\SI{6}{\kilo\metre})$ horizontal wavelength. Figure \ref{fig1}b shows the corresponding vertical velocity on a slice, with strong lee wave generation at the very rough bottom topography and propagation throughout the water column. The vertical velocities are near zero at the surface, with vertical phase lines and a modal structure in the vertical indicative of superposition of the wave field due to reflection at the surface.

This phenomenon has also been seen in other realistic simulations. \citet{deMarez2020} examined the interaction of the Gulf Stream with the Charleston Bump in high resolution realistic simulations, with a focus on lee wave generation. They found that the lee waves have a surface signature, and showed qualitative agreement with sun glitter images from satellite observations. The simulation output was compared with (unbounded) linear theory, and differences noted near the surface, where surface reflection in the simulations caused a modal structure in the vertical velocity. 

\citet{Rosso2015a} investigated topographic influence on surface submesoscales using a realistic $1/80^\circ$ resolution model of the Indian sector of the Southern Ocean, and noted surface peaks in vertical velocity (their figure 3). Lee waves reaching the surface were identified in the simulations and noted as a potential source for these increased vertical velocities, but not investigated further as the focus was on vertical velocities caused by surface submesoscales. They reasoned that enhanced near surface  vertical velocities in their figure 4d are unlikely to be generated by a lee wave evident at depth, both because the near surface vertical velocities have a vertical phase line, indicating that it is decoupled from the tilted lee wave phase lines below, and the RMS vertical velocity has a near surface maximum. However, we will show that vertical phase lines near the surface and a subsurface maximum in RMS vertical velocity are expected properties of lee waves that interact with the surface.

\citet{Bachman2017} simulated a similar region of the Drake Passage to that shown in figure \ref{fig1}a to investigate the surface submesoscale field and vertical velocities. They found regions of surface intensified RMS vertical velocity, and suggested that submesoscale circulations may not account for all such vertical velocities, with lee waves a potential source. In any case, the separation of surface submesoscales and lee waves is not clear due to their similar horizontal scales, and it is possible that they interact as a result. 

The radiation of lee waves under a changing background flow has been extensively studied in the atmospheric context, with a focus on parametrising wave drag due to isolated obstacles (mountains) in atmospheric models \citep[][and references therein]{Teixeira2014}. Particularly relevant are studies of trapped lee waves, whereby sharp changes in background flow with height allow partial wave reflection and resonance \citep{Scorer1949, Teixeira2005,Teixeira2013}, leading to high and low drag states, with clear parallels with the resonances found due to the upper boundary in the current study. In particular, \citet{Bretherton1969} performed a comprehensive linear study including a rigid lid boundary condition similar to ours. However, a rigid lid condition in the atmosphere is not realistic, so efforts were generally made to improve the treatment of the upper boundary and reduce its impact \citep{Teixeira2014}. This study is intended to demonstrate simple properties of oceanic lee waves under changing background conditions typical of the ocean, with a particular focus on their structure in the upper ocean due to the boundary condition at the surface. Typically, atmospheric lee wave studies focus on drag. In the oceanic context, both lee wave drag and mixing are important, thus our focus is  also somewhat different to the aforementioned atmospheric studies. 

The structure of this paper is as follows. In \S \ref{sec:lineartheory}, we review and derive the linear lee wave theory with viscous and diffusive terms and discuss boundary conditions, energetics, time dependence, and complications associated with the bounded solution and non-uniform background fields including resonance and critical levels. In \S \ref{sec:numsol}, we present the numerical solver in a bounded and unbounded domain and describe the modelling set-up. We present results from the linear solver in \S \ref{sec:results}, and discuss conclusions in \S \ref{sec:discussion}. 

\section{Theoretical Framework} \label{sec:lineartheory}
Following \cite{Bell1975}, we start from the rotating, incompressible, Boussinesq equations with the inclusion of Laplacian viscosity $\mathcal{A}$ and diffusivity $\mathcal{D}$:
\begin{align}
\mathbf{u}^\dagger_t + \mathbf{u}^\dagger \cdot \nabla \mathbf{u}^\dagger + \mathbf{f} \times \mathbf{u}^\dagger &= -\rho_0^{-1}\nabla p^\dagger + b^\dagger\mathbf{\hat{z}} + \mathcal{A}\nabla^2 \mathbf{u}^\dagger\,, \label{mom} \\ 
b^\dagger_t + \mathbf{u}^\dagger \cdot \nabla b^\dagger &= \mathcal{D} \nabla^2 b^\dagger\,, \label{buoy} \\
\nabla \cdot \mathbf{u}^\dagger &= 0\,, \label{incompr}
\end{align}
where $\mathbf{u}^\dagger = (u^\dagger,v^\dagger,w^\dagger)$ is the velocity, $\mathbf{f} = (0,0,f)$ is the Coriolis parameter, $p^\dagger$ is the pressure, $b^\dagger = -\rho^\dagger g/\rho_0$ is the buoyancy, $\rho^\dagger$ is the density, $\rho_0$ is a constant reference density, and $\dagger$ is used to denote total fields.
\subsection{Base state} \label{sec:basestate}
We specify that the background velocity is in the $x$-direction, and both background velocity and stratification are steady and vary only in the vertical, so that the base state is given by $\mathbf{u}^\dagger = (U(z),0,0)$, $p^\dagger = \overline{p}(y,z)$, $b^\dagger = \overline{b}(y,z)$. Assuming that the impact of perturbations on the the mean flow is not leading order, from \eqref{mom} it must satisfy both geostrophic and hydrostatic balance:
\begin{align}
-fU &= -\rho_0^{-1}\overline{p}_y\,, \label{geostroph} \\
0 &= -\rho_0^{-1}\overline{p}_z + \overline{b}\,. \label{hyd}
\end{align}
Eliminating $\overline{p}$ from \eqref{geostroph} - \eqref{hyd} gives the thermal wind balance:
\begin{equation}
-fU_z = \overline{b}_y\,. \label{twb}
\end{equation}
Requiring that the stratification $N^2 = \overline{b}_z$ is a function of $z$ only, \eqref{twb} gives that 
$fU_{zz} = 0$. We therefore only consider base states such that $fU_{zz} = 0$, but continue the derivation for general $U(z)$ for use when $f=0$. This ensures that although $\overline{p}$ and $\overline{b}$ are functions of $y$, $\overline{p}_y$ and  $\overline{b}_y$ are not, and the problem remains effectively 2D so that all coefficients of the linearised problem to be derived in \S \ref{sec:linearisation} are independent of $y$.

\subsection{Energy loss} \label{sec:energy_loss}
A representation of lee wave energy loss is crucial to understanding the structure of the lee wave field in the vertical. Lee wave energy must either be reabsorbed by the mean flow, or lost to dissipation and mixing. The latter is a result of energy transfer to smaller scales through instabilities of the waves themselves, or through nonlinear interactions with other waves and the background flow. In our idealised linear model, we cannot properly represent either the dynamics of the waves which can lead to instabilities and breaking, or small scales from other sources of turbulence that act to eventually dissipate even linear waves. The effect of this energy lost from the lee wave field must therefore be parametrised.

Parametrisation of dissipation and mixing at the sub-gridscale in models is generally implemented through Laplacian (or higher order) viscous and diffusive terms in the momentum and buoyancy equations - as shown in \eqref{mom} - \eqref{buoy}. \citet{Shakespeare2017a} provide a comprehensive overview of the role of Laplacian viscosity and diffusivity in the linear lee wave problem, with a focus on preventing excessive dissipation in wave resolving models. Here, we do not represent the processes that drain energy from the lee wave field, so aim to model them diffusively with this parametrisation. However, unlike \citet{Shakespeare2017a}, we are dealing with background flows that vary in the vertical, and thus including the vertical components $\mathcal{A}_v\mathbf{u}_{zz}^\dagger$ and $\mathcal{D}_vb_{zz}^\dagger$ of the Laplacian terms in our study significantly complicates the solution.

For mathematical convenience, we therefore represent the total viscous and diffusive terms by the horizontal components only. This allows some scale selection for energy loss (improving on, say, a simple Rayleigh friction), without overly complicating the problem. Using only the horizontal component as a proxy for the total dissipation and mixing has certain drawbacks, including invalidating any solutions where the vertical wavelength changes drastically or becomes very small, e.g. at critical levels. It is important to keep in mind the simplifications made here when analysing the model mixing and dissipation in \S \ref{sec:results}. Direct comparisons between our horizontal turbulent viscosity $\mathcal{A}_h$ and diffusivity $\mathcal{D}_h$ parameters and other studies or models should also be made with care, since they represent both horizontal and vertical viscosity and diffusivity. Furthermore, since $\mathcal{A}_h$ and $\mathcal{D}_h$ represent both background turbulent processes and breaking of the lee wave field itself, their `real' values should depend on nonlinearity of the wave field and properties of the background flow, among other things. Although the simplifications made with this parametrisation are likely to modify our solutions somewhat, we believe that the key results of this study are unaffected. 

\subsection{Linearisation} \label{sec:linearisation}
For $Fr_L \ll 1$, we consider small perturbations to the base state described in \S \ref{sec:basestate}. The coefficients of the linearised equations are independent of $y$ due to the constraints on the base state described in \S \ref{sec:basestate}, thus the perturbation variables are also taken to be independent of $y$. We also assume here that the perturbations are steady, although this need not be imposed at this point and follows from the application of the steady boundary conditions to be described in \S \ref{sec:BC}. 

Letting $\mathbf{u}^\dagger = (U(z) + u(x,z),v(x,z),w(x,z))$,  $b^\dagger = \overline{b}(y,z) + b(x,z)$, $p^\dagger = \overline{p}(y,z) + p(x,z)$ and linearising  \eqref{mom} - \eqref{incompr} gives:
\begin{align}
wU_z + Uu_x -fv &= -\rho_0^{-1}p_x + \mathcal{A}_h u_{xx}\,, \label{mom1} \\
Uv_x + fu &=   \mathcal{A}_h v_{xx}\,,  \label{mom2} \\
\alpha Uw_x &= -\rho_0^{-1}p_z + b + \alpha \mathcal{A}_h w_{xx}\,, \label{mom3} \\
Ub_x  - fvU_z +wN^2  &= \mathcal{D}_h b_{xx}\,, \label{buoy2} \\
u_{x} + w_{z} &= 0\,, \label{incompr2}
\end{align}
where $\alpha \in \{0,1\}$, so that when $\alpha = 0$ the equations are hydrostatic. The hydrostatic assumption is made when the ratio of vertical to horizontal scales is small, as is often the case for lee waves. We introduce a perturbation streamfunction $\psi$ such that $u = -\psi_z$, $w = \psi_x$, with Fourier transform $\hat{\psi}(k,z)$ defined such that:
\begin{equation}
\psi(x,z) =\frac{1}{2\pi} \int_{-\infty}^\infty \hat{\psi}(k,z)e^{ikx} dk \,. \label{ft}
\end{equation}
Taking the Fourier transform in $x$ of \eqref{mom1} - \eqref{incompr2} and solving for  the transformed streamfunction $\hat{\psi}(k,z)$ gives a second order ordinary differential equation: 
\begin{equation}
\hat{\psi}_{zz} + P(k,z)\hat{\psi}_z + Q(k,z)\hat{\psi} = 0\,, \label{main}
\end{equation}
where 
\begin{align}
P(k,z) &= \frac{f^2U_z\left(2U - ik(\mathcal{A}_h + \mathcal{D}_h)\right)}{\left(k^2(U - ik\mathcal{A}_h)^2 - f^2\right)\left(U -ik\mathcal{A}_h\right)\left(U-ik\mathcal{D}_h\right)}\,, \label{P}\\
Q(k,z) &= \frac{k^2\left(U-ik\mathcal{A}_h\right)\left(N^2 - \alpha k^2(U-ik\mathcal{A}_h)(U-ik\mathcal{D}_h)\right)}{\left(U - ik\mathcal{D}_h\right)\left(k^2(U-ik\mathcal{A}_h)^2 - f^2\right)} - \frac{k^2U_{zz}(U-ik\mathcal{A}_h)}{k^2(U-ik\mathcal{A}_h)^2 - f^2}\,. \label{Q}
\end{align}
With constant background velocity and stratification and in the absence of viscosity and diffusivity, this reduces to the familiar equation for the steady lee wave problem \citep{Bell1975}:
\begin{equation} 
\hat{\psi}_{zz}(k,z) + k^2\frac{N^2 - \alpha U^2k^2}{U^2k^2 - f^2}\hat{\psi}(k,z) = 0\,, \label{simplecase}
\end{equation}
with solution:
\begin{equation}
\hat{\psi}(k,z) = A(k)e^{im(k)z} + B(k)e^{-im(k)z}\,, \label{simplesol}
\end{equation}
for some functions $A$ and $B$ to be specified by the boundary conditions, where
\begin{equation}
m^2(k) = k^2\frac{N^2 - \alpha U^2k^2}{U^2k^2 - f^2}\,. \label{m}
\end{equation}
It is clear from \eqref{simplesol} and \eqref{m} that there are radiating solutions (lee waves) only when $m$ is real, that is when the topographic wavelength $k$ satisfies
\begin{equation}
|f| < |Uk| < |N|\,.
\end{equation}
For wavenumbers $k$ in this radiating range, rotation can be neglected when $|f| \ll |Uk|$, and the hydrostatic assumption ($\alpha = 0$) can be made when $|Uk| \ll |N|$, since in this case the vertical wavenumber $m \sim \frac{N}{U}$ (from \eqref{m}), so $|Uk|/|N|$ represents the ratio of vertical to horizontal wavelengths. 
\subsection{Boundary conditions} \label{sec:BC}
\subsubsection{Bottom boundary condition}
For a given $k$, \eqref{main} requires two boundary conditions. A free slip condition to ensure that the flow is parallel to the 2D topography $h(x)$ is given by:
\begin{equation}
w^\dagger(x,h(x)) = u^\dagger(x,h)h_x\,.
\end{equation}
Linearising about the base state then gives:
\begin{equation}
w(x,0) = U(0)h_x\,, \label{bottombc}
\end{equation}
or equivalently, defining the Fourier transform of the topography $\hat{h}(k)$ similarly to \eqref{ft}:
\begin{equation}
\hat{\psi}(k,0) = U(0)\hat{h}(k)\,.
\end{equation}
Given this requirement, we write $\hat{\psi}(k,z) = U(0)\hat{h}(k)\hat{\zeta}(k,z)$, where $\hat{\zeta}(k,z)$ is the normalised vertical structure function for a wavenumber $k$, so that
\begin{equation}
\psi(x,z) =\frac{U(0)}{2\pi} \int_{-\infty}^\infty \hat{\zeta}(k,z)\hat{h}(k)e^{ikx} dk\,, \label{psiint}
\end{equation}
and $\hat{\zeta}(k,z)$ satisfies
\begin{align}
\hat{\zeta}_{zz} + P(k,z)\hat{\zeta}_z + Q(k,z)\hat{\zeta} &= 0\,,  \label{zetamain} \\
\hat{\zeta}(k,0) &= 1\,. \label{zetaBC1}
\end{align}
\subsubsection{Upper radiating boundary condition}
For the second condition, consider first the classical unbounded lee wave problem, which requires that waves propagate freely through the upper boundary. For the uniform background and inviscid case with solution given by \eqref{simplesol}, the coefficients $P$ and $Q$ are constant in $z$, and a vertical wavenumber $m(k)$ \eqref{m} can be found. For each $k$, there is a well defined vertical group velocity (to be discussed further in \S \ref{sec:groupvel}), which must be positive when the solutions are wavelike (when $m$ is real) to ensure that energy radiates away from topography. This is ensured by choosing $m$ to have the same sign as $Uk$ when $m$ is real. When $m$ is imaginary, physical intuition necessitates that the positive root is taken so that  disturbances decay away from topography rather than increase exponentially. 

If viscosity and diffusivity are non-zero, the solution can still be found with this upper boundary condition since $P$ and $Q$ remain constant in $z$ and there is still a well defined vertical wavelength $m(k)$ (up to a sign). However, since $m$ is now complex, the correct choice for the sign of $m$ must always be that with positive imaginary part so that the solution decays away from the topography. For the weakly viscous and diffusive case such that the vertical decay scale due to viscosity and diffusivity is much greater than the vertical wavelength, this is the same choice as requiring the real part of vertical group velocity to be positive, except when $|Uk| \rightarrow |f|$, where the distinction between radiating and non-radiating solutions becomes less clear than in the inviscid case - see \citet{Shakespeare2017a} for a detailed discussion. Here, we consider the effect of weak viscosity and diffusivity on radiating lee waves from topography such that $|f| < |Uk| < |N|$. 

When the coefficients $P$ and $Q$ are not constant in $z$, this radiating upper boundary condition is poorly defined, since for each $k$ there is not a well defined vertical wavelength and group velocity. Waves can internally reflect and refract from changes in background density or velocity, so the solution cannot be restricted to upward propagating components. 

\subsubsection{Upper free surface  boundary condition}
If lee waves reach the upper ocean, the radiating upper boundary condition is inappropriate, and the air-sea interface may instead be better represented by a free surface boundary condition.  A simpler condition is the rigid lid - we will show that for this problem, these are essentially equivalent. 

At a free surface given by $z = H + \eta(x)$, where $\eta \ll H$, the linearised kinematic boundary condition (c.f. \eqref{bottombc}) is:
\begin{equation} \label{kinematicBC}
\psi(x,H) = U(H)\eta(x)\,.
\end{equation}
A further dynamic boundary condition is required to close the problem, given by $p^\dagger(x,H + \eta(x)) = p_A$, where $p^\dagger(x,z) = \overline{p}(z) + p(x,z)$ is the total pressure in the fluid, $p_A$ is the atmospheric pressure (assumed constant), and $\overline{p}(H)= p_A$. Expanding $p^\dagger(x,H + \eta(x))$ to first order in the perturbation variables and $\eta$ gives:
\begin{align}
p^\dagger(x,H + \eta(x)) &= p^\dagger(x,H) + \eta(x)\diffp{p^\dagger}{z}\bigg{|}_{z=H}\,, \\
&= p_A + p(x,H) + \eta(x)\diffp{\overline{p}}{z}\bigg{|}_{z=H}  \,.
\end{align}
Using hydrostatic balance of the base state \eqref{hyd} then gives the dynamic boundary condition:
\begin{equation} \label{dynamicBC}
p(x,H) = -\rho_0 \overline{b}(H)\eta(x) = \rho_0 g \eta(x) \,,
\end{equation}
where the reference density $\rho_0$ is taken to be the base state surface density. 

Eliminating the unknown $\eta$ from the surface boundary conditions \eqref{kinematicBC} and \eqref{dynamicBC} gives the boundary condition:
\begin{equation} \label{surfBC}
\psi(x,H) = \frac{U(H)p(x,H)}{\rho_0 g}\,.
\end{equation}
This surface boundary condition could be used with \eqref{bottombc} to solve \eqref{mom1} - \eqref{incompr2} , then the surface height recovered from \eqref{dynamicBC} or \eqref{kinematicBC}. However, in practise this is unnecessary if the surface height is not of interest, as \eqref{surfBC} can be well approximated by the rigid lid condition $\psi(x,H) = 0$, equivalent to imposing $\eta(x) = 0$ (and thereby not satisfying the dynamic boundary condition). To see why, first notice from \eqref{mom1} that for negligible rotation, shear, and viscosity, $p \sim -\rho_0 U u \sim \rho_0 U \psi_z$. For slowly varying background conditions, we expect $\psi(x,z)$ to locally have a sinusoidal structure in the vertical, so let (for fixed $x$):
\begin{align}
\psi(z) &\sim A\sin (mz + \varphi)\,, \\
\psi_z(z) &\sim Am\cos (mz + \varphi)\,,
\end{align}
for some amplitude $A$, wavenumber $m$, and phase $\varphi$. Then, using the boundary relation \eqref{surfBC}:
\begin{equation}
\sin( mH + \varphi) \sim \frac{mU^2}{g}\cos (mH + \varphi)\,.
\end{equation}
Assuming that $m \sim N/U$ (the hydrostatic, non-rotating limit of \eqref{m}):
\begin{equation}
\tan (mH + \varphi) \sim \frac{NU}{g} \ll 1\,,
\end{equation}
even for large upper ocean values of $U$ and $N$, and this scaling still holds for realistic conditions with rotation and nonhydrostatic waves. Therefore, the phase $\varphi$ is such that $\psi(H) \simeq 0$, and it is clear that a rigid lid approximation is sufficient for determining the interior structure of the lee waves. The full free surface boundary condition could be implemented to determine exactly the (linear) surface height $\eta(x)$, but hereafter we only consider the rigid lid boundary condition. Since the interior flow is relatively unaffected by this approximation, we could still estimate the surface height without explicitly solving for it, using \eqref{dynamicBC}:
\begin{equation}
\eta(x) \sim \frac{p(x,H)}{\rho_0 g}\,,
\end{equation}
where $p(x,H)$ is found from the rigid lid solution.
\subsubsection{Upper rigid lid  boundary condition}
With the rigid lid condition $\psi(x,H) = 0$, the solution to the bounded problem is then given by \eqref{zetamain} - \eqref{zetaBC1}, with the upper boundary condition:
\begin{equation}
\hat{\zeta}(k,H) = 0\,. \label{zetaBC2}
\end{equation}
\subsection{Group velocities} \label{sec:groupvel}
The behaviour of lee waves in a bounded domain depends strongly on the direction of their group velocity. Consider the inviscid and unbounded problem with uniform background stratification and velocity, so that the vertical wavenumber $m$ is independent of $z$. Re-deriving the governing equation \eqref{simplecase} with time dependence by considering plane wave solutions $\sim e^{i(kx + mz -\omega t)}$ gives the dispersion relation (c.f. \eqref{m}):
\begin{equation} \label{disprel}
(\omega - Uk)^2 = \frac{N^2k^2 + f^2m^2}{\alpha k^2 + m^2}\,,
\end{equation}
where $\omega = 0$ for steady lee waves satisfying the boundary condition \eqref{bottombc}. The phase velocity is zero as a result, but the group velocity is non zero and can be found by differentiating \eqref{disprel}:
\begin{align}
\mathbf{c_g} &= \left(\diffp{\omega}{k}, \diffp{\omega}{m}\right)\,, \\
&= \left( \frac{f^2(N^2 - \alpha U^2k^2) + \alpha U^2k^2(U^2k^2 - f^2)}{Uk^2(N^2 - \alpha f^2)} , \frac{(U^2k^2 - f^2)^\frac 32(N^2 - \alpha U^2k^2)^\frac 12 }{Uk^2(N^2-\alpha f^2)}\right)\,, \label{groupvel}
\end{align}
where the sign of the vertical group velocity is taken to be positive when $m$ is real, as is appropriate for the unbounded case, and it is assumed that $|f| < |Uk| < |N|$ so that the waves are radiating. 
\begin{figure} 
\centering
\includegraphics[width=5.3in]{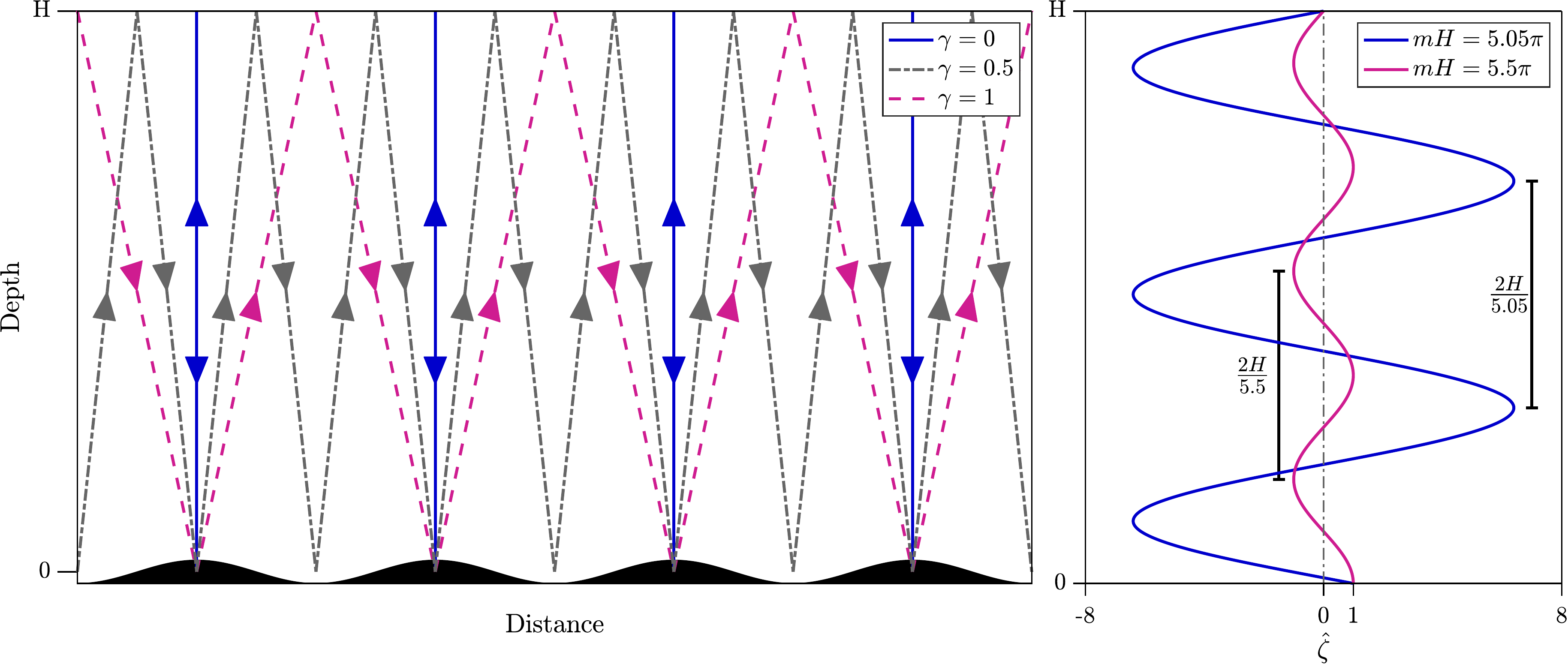}  \\
\caption{a) Diagram showing monochromatic topography with indicative ray paths for several values of the overlap parameter $\gamma$, demonstrating some possible idealised paths of lee waves with different directions of group velocity. b) Diagram showing the vertical structure function $\hat{\zeta}(k,z)$ for the analytic solution \eqref{analyticsol}, for some vertical wavenumbers $m$ such that the solution is near-resonant (blue) and at its minimum amplitude (pink). }\label{fig2} 
\end{figure}
It is clear from \eqref{groupvel} that in the non-rotating and hydrostatic case ($f = \alpha = 0$), the horizontal component of group velocity is zero, and waves propagate vertically upwards. Supposing now that they encounter the surface, the waves will reflect and propagate directly downwards - still with zero horizontal group velocity and now with negative vertical group velocity - superimposing exactly on the upward propagating wave field. This scenario is illustrated for monochromatic topography in figure \ref{fig2}a (blue lines). The reflected waves can then be expected to directly increase or decrease the topographic wave drag and energy conversion by constructive or destructive interference with the upwards propagating wave field at the topography. The extent to which this occurs is determined by the energy lost to mixing and dissipation during propagation, to be discussed in \S \ref{sec:results1}. When the horizontal group velocity is zero, no energy propagates downstream, so without dissipative energy loss there can be no energy conversion into lee waves at the topography and also no wave drag. However, there may be resonance (to be discussed in \S \ref{sec:resonance}). 

If $|Uk|$ is of comparable magnitude to the Coriolis or buoyancy frequency, the waves will have a positive horizontal component of group velocity and will propagate both upwards and downstream, reflecting at the surface downstream of the topography. Without dissipation and mixing this could continue indefinitely and allow the lee wave energy to propagate far downstream, although in reality it seems unlikely that a significant amount of wave energy would undergo multiple reflections due to nonlinear interactions near the bottom boundary. For an isolated topographic peak (which will generate a continuous range of wavenumbers $k$), if the angle of propagation is large enough the reflected wave will not significantly interact with the generation process and the wave drag will be unchanged from the unbounded case. If the bump is not isolated, the reflected wave could be incident on the generation of a lee wave at different topographic feature, and the drag (and energy flux) modification would be more complex.

\begin{figure} 
\centering
\includegraphics[width=5.3in]{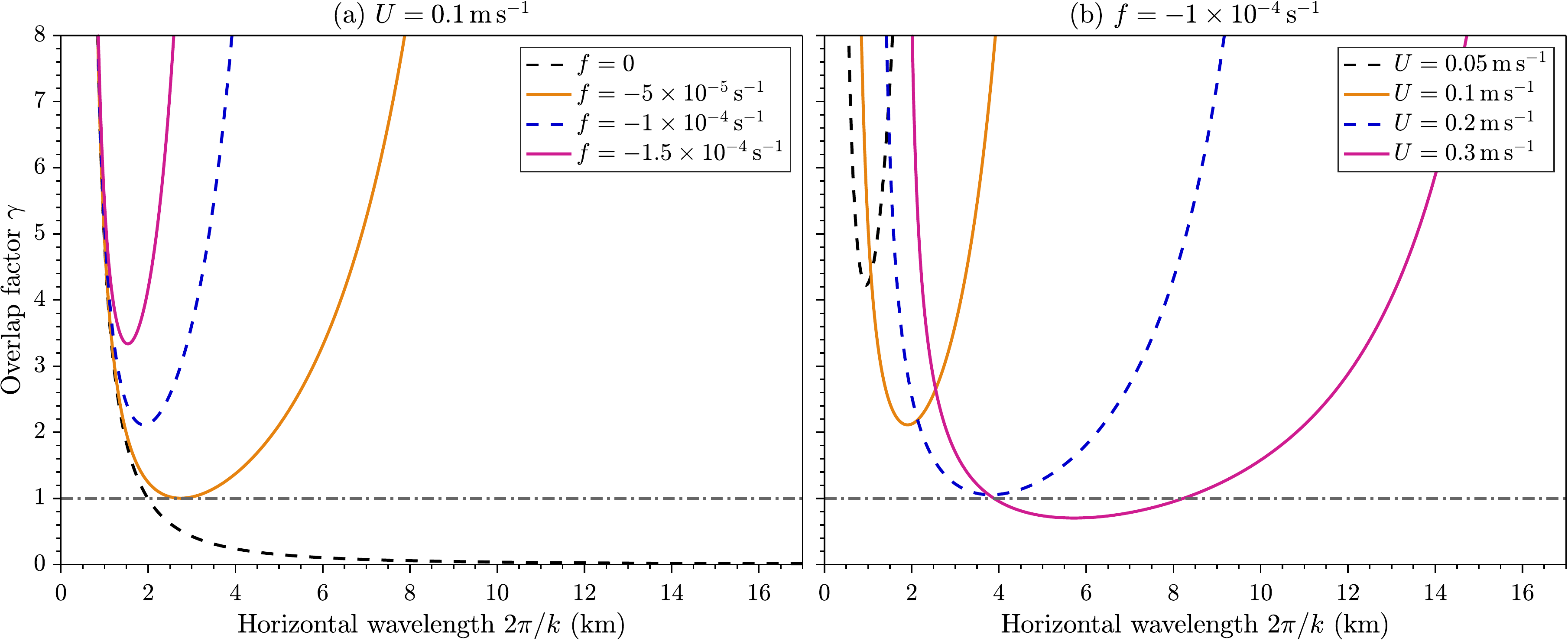}  \\
\caption{Overlap parameter $\gamma$ defined in \eqref{overlap} for (a) fixed $U = \SI{0.1}{\metre\per\second}$, varying $f$ and (b) fixed $f = \SI{-1e-4}{\per\second}$, varying $U$, with $H = \SI{3}{\kilo\metre}$, $N = \SI{1e-3}{\per\second}$. The black dashed line shows $\gamma = 1$, below which a reflected lee wave could be expected to modify its own generation mechanism.}\label{fig3} 
\end{figure}

To determine the likelihood of a lee wave superimposing on itself at the topography, we can determine the angle of propagation using \eqref{groupvel}, assuming for simplicity that $U$ and $N$ are constant and viscosity and diffusivity are negligible. An `overlap parameter' $\gamma (k)$ can be defined as:
\begin{equation} \label{overlap}
\gamma (k) = \left|\frac{kH}{\pi}\right|\tan \theta \,, \hspace{1cm} \tan \theta = \diffp{\omega}{k} \bigg/ \diffp{\omega}{m}\,,
\end{equation}
where $\theta$ is the angle of wave propagation to the vertical, and for each $k$, $\gamma$ is the horizontal distance travelled by a wave whilst propagating to the surface at $z=H$ and back to the topography at $z=0$, normalised by the horizontal wavelength. This is illustrated for $\gamma = 0, 0.5$, and $1$ in figure \ref{fig2}a. Figure \ref{fig3} shows the variation in $\gamma$ with horizontal wavelength $2\pi/k$ for varying $U$ and $f$. Each curve tends to infinity (not shown) at $k = |N/U|$ and $k = |f/U|$, at which point the vertical group velocity reaches zero and the solutions become evanescent. The increase in $\gamma$ for small horizontal wavelengths is due to the increasing downstream component of group velocity when the waves are nonhydrostatic, and for large horizontal wavelengths, due to rotation.

For smaller values of $f$ and larger values of $U$, there exists a range of scales at which $\gamma \lesssim 1$, indicating that reflected lee waves could impact on the generation mechanism by direct superposition. For $f = \SI{-1e-4}{\per\second}$, characteristic of the Southern Ocean, there exist horizontal scales at which this may be the case for $U \gtrsim \SI{0.2}{\metre\per\second}$. However, for $f = \SI{-1e-4}{\per\second}$ and $U = \SI{0.1}{\metre\per\second}$ (orange line), $\gamma > 2$ for all radiating wavelengths, and all reflected waves return to topography at least $\SI{4}{\kilo\metre}$ downstream of the generating topographic feature. Of course, this argument doesn't cover the more likely scenario of varying velocity and stratification with height. We conclude that for lee waves in shallow areas, low latitudes, or high background flows it is possible for lee waves generated by isolated topography to reflect at the surface and modify the original wave drag and energy conversion, but that this is unlikely for deep generation, low background velocities, and high latitudes. 

When the topography is not isolated, and in particular when an artificially discrete topographic spectrum is used as in this study, the wave drag modification can be significant even when the overlap parameter is larger than one. For monochromatic topography, when $\gamma(k)  = n \in \mathbb{N}$, a wave generated at a topographic peak reflects at the surface and is incident on the topographic peak $n$ wavelengths downstream from the original, as shown in figure \ref{fig2}b for $n=1$, and impacts the wave field at that generation site in a similar way to the case $\gamma = 0$.

\subsection{Analytic solution} \label{sec:analyticsol}
When $U$ and $N$ are constant with height, so that $P$ vanishes and $Q$ is a function of $k$ only, the solution for $\hat{\zeta}(k,z)$ is (extended from \citet{Baines1995}):
\begin{equation} \label{analyticsol}
\hat{\zeta}(k,z) = \frac{\sin(m(k)(H-z))}{\sin ( m(k)H )}\,,
\end{equation}
where $m(k)$ is the complex vertical wavenumber defined by $m^2(k) = Q(k)$, and the choice of sign does not matter. The solution can then be found numerically for general topography via \eqref{psiint}, or analytically for monochromatic topography $h(x) = h_0\cos k_0 x$ to be:
\begin{equation} \label{monochrom_sol}
\psi(x,z) = Uh_0 \Re \left(\frac{\sin(m(k_0)(H-z))}{\sin ( m(k_0)H )}e^{ik_0 x}\right)\,.
\end{equation}
The above solutions are valid only when $|m(k)H| \neq n\pi$, $n \in \mathbb{N}$. At such points, resonances of the system occur.

\subsection{Resonance} \label{sec:resonance}

Under the assumption that lee waves are hydrostatic ($|Uk| \ll |N|$), rotation is unimportant ($|Uk| \gg |f|$), and the system is inviscid, the vertical wavenumber is simply $m(k) = N/U$. The resonances of \eqref{analyticsol} are then independent of $k$, and occur when $|NH/U| = n\pi$ for some $n \in \mathbb{N}$. There are no steady solutions to \eqref{zetamain}, \eqref{zetaBC1} and \eqref{zetaBC2} if this condition is met. Physically, this occurs when a whole number of half-wavelengths fit in the vertical domain and there is constructive interference of the upwards and downwards propagating waves. Figure \ref{fig2}b shows the vertical structure function $\hat{\zeta}$, defined in \eqref{analyticsol}, for two real values of $m$. When $mH = 5.05\pi$ (blue) the system is near resonance, as the half-wavelength nearly divides the depth $H$ (true resonance is at $mH = 5\pi$). Thus, $\hat{\zeta}(z) = 0$ near $z=0$, so in order to satisfy the boundary condition $\hat{\zeta}(z=0) = 1$, the amplitude of the wave must be very large. At true resonance, this boundary condition cannot be met. In the opposite case, (shown for $mH = 5.5\pi$ in pink), there is destructive interference and the amplitude is at a minimum.

Under the above assumptions, the horizontal group velocity is zero, therefore energy cannot escape downstream and the wave generation at resonance continually reinforces the wave field. If this were to happen in practise, the wave amplitude would become large enough to invalidate the linearity of the wave field, perhaps causing nonlinear wave breaking or modifying the wave field or the boundary condition so as to move the system away from resonance. 

When nonhydrostaticity is included, the horizontal group velocity is non-zero and the nature of the resonance changes slightly. The vertical wavenumber $m(k) = \sqrt{N^2/U^2 - k^2}$, thus the solution \eqref{analyticsol} has singularities at
\begin{equation}
k^2 = \frac{N^2}{U^2} - \frac{n^2\pi^2}{H^2}\,, \hspace{1cm} n \in  \mathbb{N}\,. \label{poles}
\end{equation}

Physically, these singularities still represent modes where an exact number of half vertical wavelengths fit in the domain, but now this happens at different values of $U$, $N$ and $H$ for each component $k$ of the wave field. 

Mathematically, the resulting singularities of \eqref{analyticsol} are simple poles, so when the topographic spectrum $\hat{h}(k)$ is continuous (as for isolated topography), the integral \eqref{psiint} along the real line can be moved to a contour of integration in complex $k$ space that avoids the poles. To ensure that there is no disturbance at upstream infinity, the contour must be taken below rather than above the poles \citep{McIntyre1972}. The solution can then be expressed as the Cauchy principle value of \eqref{psiint} plus half the residues of the simple poles, which represent the nonhydrostatic resonant modes \citep{Baines1995}. The solutions, when $\hat{h}(k)$ is continuous, could be found numerically from \eqref{psiint} by choosing some contour of integration sufficiently far from the poles to avoid numerical difficulties. However, this becomes more difficult once rotation is included since the poles no longer all lie on the real axis. The numerical solution is also problematic since periodicity in the horizontal is assumed by default when taking a discretised Fourier transform, leading to spurious waves upstream of the isolated topography. If the topographic spectrum is discrete and includes one of the singular wavelengths defined by \eqref{poles}, then true resonance occurs and no steady solution exists. 

The inclusion of energy loss through viscosity and diffusivity aids the numerical solution by moving all poles off of the real line so that the integral \eqref{psiint} can be found numerically with a simple fast Fourier transform (FFT). Although true resonance is avoided, states can still be near resonant, as will be shown in \S \ref{sec:results1}. The topographic representation used here (see \S \ref{sec:topo}) consists of a spectrum of topographic wavenumbers, which numerically becomes a sum of discrete components. This is likely to enhance the resonance effect compared to a more realistic and inhomogeneous topography. 

\subsection{Energy and momentum} \label{sec:energetics}
The vertical linear lee wave energy flux at a given height is given by $\overline{pw}$, where an overbar represents a horizontal average. At the topography ($z=0$), this is equal to the bottom mean flow velocity multiplied by the horizontally averaged form drag exerted by the topography on the mean flow, since using \eqref{bottombc}:
\begin{equation} \label{formdrag}
\overline{pw}|_{z=0} = U(0)\overline{ph_x}|_{z=0}\,.
\end{equation}
Taking the inner product of \eqref{mom1} - \eqref{mom3} with the perturbation velocity and multiplying \eqref{buoy2} by the perturbation buoyancy gives the energy equation for the wave field. Taking a horizontal average and assuming a periodic domain in the horizontal then gives an expression for the divergence of the energy flux:
\begin{equation}
\overline{pw}_z = -\rho_0(U_zF + \overline{D})\,, \label{energyeqn}
\end{equation}
where $\overline{D} = \overline{\varepsilon} + \overline{\Phi}$ is the horizontally averaged energy loss from the flow, consisting of the dissipation rate $\varepsilon = \mathcal{A}_h|\mathbf{u}_x|^2$ and irreversible mixing $\Phi = \mathcal{D}_hb_x^2/N^2$,  and
\begin{equation}
F = \overline{uw} - \frac{f\overline{vb}}{N^2} \label{EPdef}
\end{equation}
is the wave pseudomomentum flux, or the Eliassen-Palm (E-P) flux \citep{Eliassen1960}.
If there are no critical levels  ($U \neq 0$) it can be shown from \eqref{mom1} - \eqref{buoy2} that the E-P flux $F$ is related to the energy flux as \citep[extended from][]{Eliassen1960}:
\begin{align}
\overline{pw} &= -\rho_0U\left[ \overline{\left(u - u_x\mathcal{A}_h/U\right)w} - \frac{f}{N^2}\overline{\left(b - \mathcal{D}_hb_x/U\right)v}\right] \\
&= -\rho_0UF\left(1 + \mathcal{O}\left(\mathcal{A}_hk/U\right)\right)\,, \label{EPenergy}
\end{align}
where $k$ is the characteristic wavenumber of the topography, and $\mathcal{A}_h \sim \mathcal{D}_h$. Taking typical values considered here, $\mathcal{A}_h  \sim \SI{1}{\metre\squared\per\second}$, $k \sim \SI{0.005}{\per\metre}$, and $U \sim \SI{0.1}{\metre\per\second}$, gives $\mathcal{A}_h k/U \sim 0.05 \ll 1$. Thus the energy flux is approximately equal to the local velocity multiplied by the E-P flux even when there is energy loss. 
In the inviscid problem, \eqref{energyeqn} and \eqref{EPenergy} together give \citep{Eliassen1960}:
\begin{equation} \label{forceonflow}
F_z =0\,.
\end{equation}
Therefore, the E-P flux is conserved when there is no energy lost to dissipation and mixing. When there is also no vertical shear of the mean flow ($U_z = 0$), \eqref{energyeqn} gives that the energy flux $\overline{pw}$ is also conserved. When the mean velocity increases or decreases with height, the energy flux increases or decreases correspondingly, but the E-P flux is still conserved. Any divergence of the E-P flux thus corresponds to the force exerted on the flow by the waves as they dissipate \citep{Andrews1976}. It is the divergence of $F$ rather than the Reynolds stress (or momentum flux) $\overline{uw}$ that gives the relevant lee wave forcing on the mean flow, since $\overline{uw}$ is in general not conserved - a paradox explained by \citet{Bretherton1969}.

The total wave drag on the mean flow is therefore given by the integral of $\rho_0 F_z$ over the depth of the ocean. Since there cannot be any energy or momentum flux through the upper boundary, $\overline{pw}|_{z = H} = F(H) = 0$, thus the wave drag is given by $-\rho_0 F(0)$. Comparison of \eqref{formdrag} and \eqref{EPenergy} then shows that up to $\mathcal{O}(\mathcal{A}_h k/U)$ the wave drag is equal to the form drag.

Since $\overline{pw}|_{z = H} = F(H) = 0$, if energy loss $\overline{D}$ is zero, $F = 0$ everywhere (from \eqref{forceonflow}) and $\overline{pw} = 0$ everywhere (from \eqref{EPenergy}), thus there is no topographic wave drag on the flow or energy conversion to lee waves in steady state. Energy loss is therefore a key component in the bounded study, as there can be no topographic wave drag without it. Of course, in the unbounded problem there must also be energy loss in order to have wave drag at the topography - but that energy loss can implicitly occur by allowing the lee waves to exit the given domain (such that $\overline{pw}|_{z =H} > 0$) and dissipate `elsewhere'. 

From \eqref{energyeqn}, the wave energy flux can change both by exchange with a mean flow through the E-P flux \citep{Kunze2019}, and by mixing and dissipation. Integrating \eqref{energyeqn} over the entire height of the domain gives:
\begin{equation}
\overline{pw}|_{z=0} -\overline{pw}|_{z=H} = \rho_0\int_0^H U_zF + D\, dz\,. \label{energyint}
\end{equation}

If there is an upper boundary and no background shear ($U_z = 0$), then $\overline{pw}|_{z=H} = 0$, and topographic energy conversion and wave drag are directly proportional to the total mixing and dissipation in the water column.

\subsection{Time dependence} \label{sec:time}
When calculating lee wave fluxes, it is usually assumed that the background fields and lee waves are steady. In reality, the geostrophic flows that generate lee waves vary on timescales of days to weeks. Transient waves that are generated when the background flow changes are unaccounted for, and the time taken for the steady lee wave field to equilibriate to the solutions found by the steady solver could be long compared to the typical timescales of the flow. 

\begin{figure} 
\centering
\includegraphics[width=5.3in]{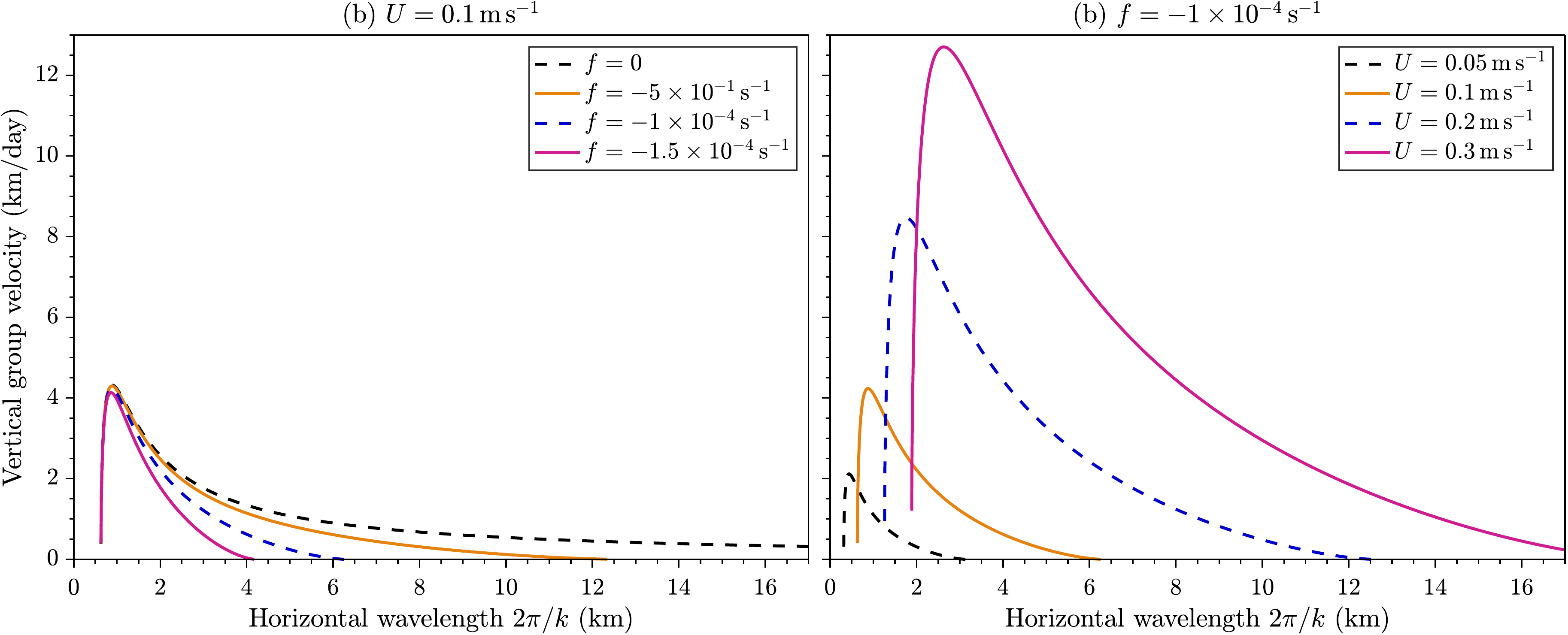} \\
\caption{Vertical group velocity defined in \eqref{groupvel} for (a) fixed $U = \SI{0.1}{\metre\per\second}$, varying $f$ and (b) fixed $f = \SI{-1e-4}{\per\second}$, varying $U$, with $H = \SI{3}{\kilo\metre}$, $N = \SI{1e-3}{\per\second}$. } \label{fig4} 
\end{figure}

The relevant timescale here is the time taken for the lee wave to propagate from the topography to the surface. Figure \ref{fig4} shows the vertical group velocity (defined in \eqref{groupvel}) for various values of $f$ and $U$. Lee waves generated at smaller horizontal scales (larger $k$) propagate faster, although they are also more likely to dissipate along the way due to sharper horizontal gradients. The effect of rotation on larger scales significantly slows the vertical group velocity, so that larger horizontal scale waves will take significantly longer to develop. The group velocity increases with $U$, so larger background velocities allow faster lee wave propagation. For a wave with wavelength $\SI{3}{\kilo\metre}$ in a background flow $U = \SI{0.1}{\metre\per\second}$, $N = \SI{1e-3}{\per\second}$ and $f = \SI{-1e-4}{\per\second}$, the vertical group velocity is approximately $\SI{1}{\kilo\metre}$ per day, suggesting that a wave would take 3 days to propagate to the surface and a further 3 days to reflect back to the topography in an ocean of depth $\SI{3000}{\metre}$. The timescale separation of full water-column lee wave formation and the mesoscale eddy field is therefore not clear, and depends on the scale of the waves. However, in energetic regions of the ocean such as the Drake Passage shown in figure \ref{fig1}, large velocities can enable high vertical group velocities and the steady approximation for lee waves at certain scales is expected to be valid. 

\subsection{Critical levels} \label{sec:critical}

In the inviscid and non-rotating problem, there are singularities of \eqref{main} - \eqref{Q} at levels where $U = 0$ \citep{Booker1967, Maslowe1986}. These are known as critical levels, where the horizontal phase speed of the wave (here equal to zero) equals the mean flow speed. At these levels, the vertical wavelength and group velocity vanish. No energy or momentum flux at the original wavenumber can propagate any further vertically, and the perturbation velocities become very large, invalidating the linear solution. In reality, instabilities and energy loss can lead to wave breaking and reflection at this level \citep{Wurtele1996}, thus critical levels may be a sink of lee wave energy in the ocean \citep{Bell1975}. However, this requires that the mean flow speed reaches zero somewhere in the water column. This is not an ubiquitous feature of the geostrophic eddies of the ACC, although critical levels may exist. This mechanism may be more important in regions of layered currents such as near the equator or in western boundary currents.  

When rotation is included, there exist two further singularities of \eqref{main} at $U= \pm |f/k|$, above and below the critical level $U = 0$ \citep{Jones1967}. These act to prevent the vertical propagation of the wavenumber $k$, in a similar way to the critical level of the non-rotating problem at $U = 0$. However, since each critical level is specific to the wavenumber $k$ (unlike for the non-rotating problem), if the spectrum of the topography is continuous it can be shown that there need not be singularities of the linear problem at these critical levels since the relevant solutions of \eqref{main} are logarithmic and thus integrable over a spectrum \citep{Wurtele1996a}. Therefore, in reality there is not a single well defined critical level for lee waves with rotation and a continuous spectrum of wavenumbers. However, the solutions may still become nonlinear so as to invalidate the linear solution and cause breaking. It can also be shown that when $f \neq 0$, the solution at $U = 0$ is no longer singular \citep{Grimshaw1975} - physically this is because all components have already reached their first critical level and stopped propagating.
 
When the flow is sheared such that $|U|$ decreases with height, energy transfers from the lee waves to the mean flow via the E-P flux (see \eqref{energyeqn}), leaving less energy to be dissipated at the critical level for a particular wavenumber $k$. \citet{Kunze2019} examine this mechanism as a possible sink for lee wave energy in regions of intensified bottom flow. In particular, for lee wave energy generated at wavenumbers far from the inertial limit $|Uk| = |f|$, a greater proportion of the initial energy is available to be reabsorbed by a mean flow decreasing with height, allowing a smaller percentage to be dissipated at the critical level at $|Uk| = |f|$ or elsewhere. For waves generated close to the inertial limit (from large scale topography), little energy is available for transfer to the mean flow, as it will instead soon reach its critical level and dissipate. 

The inclusion of viscosity and diffusivity allows non-singular solutions to be found at critical levels where $|Uk| = |f|$. However, near these levels the wave fields can become nonlinear, invalidating the linear approach. Furthermore, on the approach to these levels the vertical wavelength tends to zero, creating sharp vertical gradients and enhancing energy loss. Having neglected vertical viscosity and diffusivity in our solution, this energy loss does not take place. When the horizontal viscosity and diffusivity are large enough and shear small enough for the solutions to stay appropriately linear as a critical level is approached, the linear solution is valid, but may be unrealistic due to the lack of vertical dissipation and mixing. Velocity profiles that decrease with height are therefore not considered hereafter. 

For positively sheared background flows where the flow speed increases with height, energy instead transfers from the mean flow to the lee waves during propagation (see \eqref{energyeqn}). Since wind driven oceanic currents tend to be surface intensified, this may be a common occurrence. In this case (or if stratification $N$ decreases with height), the waves may reach `turning levels', whereby their intrinsic frequency $Uk$ reaches the buoyancy frequency $N$ \citep{Scorer1949}. At such levels the vertical wavenumber $m$ tends to zero, and the wave is reflected downward. \citet{Scorer1949} showed that wave amplitudes in the resulting `trapped' wave field could be increased by the superposition of reflected waves, much like in the current study due to the upper boundary. These turning levels are not the focus of our study, but may occur in the solutions for certain wavenumbers.

\section{Numerical solution} \label{sec:numsol}
The solution to the viscous linear lee wave problem will be found subject to both the radiating upper boundary condition (in which case we require $U$ and $N$ to be constant with height, as discussed in \S \ref{sec:BC}) and the rigid lid upper boundary condition, in which case we consider general $N^2 > 0$ and $U(z)$ such that $U > 0$, $U_z > 0$ and $fU_{zz} = 0$ (see \S \ref{sec:basestate}). 
\subsection{Unbounded solver} \label{sec:numsolUB}
The solution can be found in the traditional way \citep{Bell1975}, with the requirement the solution decays away from the topography as discussed in \S \ref{sec:BC}. The solution for $\psi$ is given by \eqref{psiint}, where $\hat{\zeta}(k,z)$ satisfies \eqref{zetamain} - \eqref{zetaBC1}, with the radiating upper boundary condition satisfied by taking the correct choice of branch for $m$. Note that $P(k,z) = 0$ and $Q(k,z) = Q(k)$, so the solution for $\hat{\zeta}$ is simply:
\begin{equation}
\hat{\zeta}(k,z) = e^{im(k)z}\,,
\end{equation}
where $m^2(k) = Q(k)$ and $\Im(m) > 0$. This can be implemented numerically for general topography $h(x)$ by performing the Fourier transforms with a FFT. Once $\psi$ is found, all other wave fields can be recovered. 
\subsection{Bounded solver} \label{sec:numsolB}
When the background flow is uniform in $z$, the solutions can be found similarly to the unbounded case above, using the analytic solution \eqref{analyticsol} for $\hat{\zeta}(k,z)$. When $U$ and $N$ are not uniform, we use Galerkin methods to solve \eqref{zetamain} - \eqref{zetaBC2}, an unforced second order ordinary differential equation with inhomogeneous boundary conditions. First, we transform it into a forced problem with homogeneous boundary conditions. Let
\begin{equation}
\hat{\zeta}(k,z) = \hat{\phi}(k,z) + G(k,z)\,,\label{etadef}
\end{equation}
where $G$ is some function such that $G(k,0) = 1$ and $G(k,H) = 0$. Then $\hat{\phi}$ satisfies
\begin{align}
\hat{\phi}_{zz} + P(k,z)\hat{\phi}_z + Q(k,z)\hat{\phi} &= R(k,z)\,, \label{phimain}\\
\hat{\phi}(k,0) &= 0\,, \label{phiBC1} \\
\hat{\phi}(k,H) &= 0 \label{phiBC2}\,,
\end{align}
and $R$ satisfies
\begin{equation}
G_{zz} + P(k,z)G_z + Q(k,z)G = - R(k,z)\,. \label{GR}
\end{equation}
$G$ can be chosen to be any function satisfying $G(k,0) = 1$ and $G(k,H) = 0$. We choose it so that $R(k,0) = R(k,H) = 0$ by taking $G$ to be a cubic polynomial in $z$, and solving for the coefficients. $R$ can then be found via \eqref{GR}. 
The problem \eqref{phimain} - \eqref{phiBC2} can now be solved numerically using Galerkin methods. Specifically, for each $k$ we decompose $\hat{\phi}$, $P$, $Q$, and $R$ into finite Fourier sums with some truncation limit $M$:
\begin{align}
\hat{\phi}(k,z) &= \sum_{m=1}^Ma_m(k)\sin \frac{m\pi z}{H}\,, \hspace{1cm} &P(k,z) &= \sum_{j=1}^Mp_j(k)\sin \frac{(j-1)\pi z}{H}\,,  \nonumber \\
Q(k,z) &= \sum_{i=1}^Mq_i(k)\cos \frac{(i-1)\pi z}{H}\,, \hspace{1cm} &R(k,z) &= \sum_{n=1}^Mr_n(k)\sin \frac{(n-1)\pi z}{H} \,, \label{expansions} 
\end{align}
where the $q_i$, $p_j$ and $r_n$ are known and found via the relevant sine or cosine transform, and the coefficients $a_m$ are to be found. Notice that the sine expansion of $\hat{\phi}$ and $R$ ensures that their boundary conditions are satisfied. However, if $P \neq 0$ or $Q_z \neq 0$ at $z = 0,H$, the sine and cosine expansions of $P$ and $Q$ respectively must represent one or more discontinuities in $P$ or $Q_z$ at endpoints. The numerical solution is therefore an approximation that is valid only in the interior, although \eqref{phimain} is satisfied everywhere by the series expansions. There can also be noise at the frequency of the truncation limit near the endpoints of the series representations due to the Gibbs phenomenon. With increasing truncation limit and vertical resolution, the interior series solution approaches the actual solution at all interior points - \eqref{energyeqn} can be used to validate this. As a consequence, quantities should not be evaluated at $z=0,H$, and \eqref{energyint} is used to find the wave drag rather than direct evaluation at $z=0$. 

Substituting \eqref{expansions} into \eqref{phimain}, integrating over $z \in [0,H]$ and using the orthogonality properties of sine and cosine gives a matrix equation for the coefficients $a_m$:
\begin{equation}
A_{mn}a_n = B_{mn}r_n\,, \label{mateqn}
\end{equation}
where:
\begin{align}
A_{mn} &= -\left(\frac{n\pi}{H}\right)^2\delta_{m,n} + \frac{n\pi}{2H}(p_{m-n+1} + p_{m+n+1} - p_{n-m+1}) + \frac{1}{2}(q_{m-n+1} + q_{n-m+1} - q_{m+n+1})\,,\\
B_{mn} &= \delta_{n,m+1}\,.
\end{align}
The $a_m$ can now be found from \eqref{mateqn} by inverting the matrix $A$. $\hat{\phi}$ can then be recovered from the coefficients $a_m$, $\hat{\zeta}$ found from \eqref{etadef}, and $\hat{\psi}$ found from \eqref{psiint}. 

\subsection{Topography} \label{sec:topo}
The topography to be used with the numerical solver is similar to that used in previous lee wave modelling studies \citep{Nikurashin2010b,Nikurashin2011a,Nikurashin2014,Klymak2018,Zheng2019}, found from the theoretical abyssal hill topographic spectrum $P_{2D}(k,l)$ proposed by \citet{Goff1988}: 
\begin{equation}
P_{2D}(k,l) = \frac{2\pi h_0^2 (\mu - 2)}{k_0 l_0}\left(1 + \frac{k^2}{k_0^2} + \frac{l^2}{l_0^2}\right)^{-\frac{\mu}{2}}\,, 
\end{equation}
by integrating over wavenumbers $l$. $k_0$ and $l_0$ are the characteristic horizontal wavenumbers, $\mu$ is the high wavenumber spectral slope, and $h_0$ is the RMS abyssal hill height. For comparison with other recent lee wave studies, we set $k_0 = \SI{2.3e-4}{\per\metre}$, $l_0 = \SI{1.3e-4}{\per\metre}$ and $\mu = 3.5$, in line with representative parameters of the Drake Passage region used in \citet{Nikurashin2010a, Zheng2019}. Next, $P_{1D}(k)$ is set to zero for wavenumbers $k$ such that $|U(0)k| < |f|$ or $|U(0)k| > |N(0)|$, since solutions in these ranges are non-propagating. The typical values used are $N(0) = \SI{1e-3}{\per\second}$, $U(0) = \SI{0.1}{\metre\per\second}$ and $f = \SI{-1e-4}{\per\second}$, corresponding to a topography with wavelengths between $\sim \SI{630}{\metre}$ and $\sim \SI{6300}{\metre}$. Note that the same topography is used throughout, even when $f =0$. 

The topographic height differs from that used in the aforementioned studies, since the solver is linear and the solutions must therefore remain approximately linear to be valid. We normalise the topography resulting from the above steps so that the RMS of the final topography $h_{rms} = \SI{25}{\metre}$. This gives a Froude number $Fr_L = Nh_{rms}/U = 0.25$ and is sufficient to keep the solution near linear such that the perturbation horizontal velocity $u$ is less than the background velocity $U$, with the exception of resonant cases. This is an unrealistically low Froude number for the rough topography of many parts of the Southern Ocean \citep{Nikurashin2010a}, but since the perturbation quantities are linear in $\hat{h}(k)$ (e.g. \eqref{psiint}), simple scaling arguments can recover the dependence on $h_{rms}$. The goal of this study is not to make predictions of the actual magnitude of the lee wave field, but its structure in the vertical and dependence on viscosity and diffusivity, background fields, and boundary condition. 

\subsection{Numerical set-up} \label{sec:setup}

In the following section, the numerical solver is used to solve for the wave fields in a 2D domain of width $\SI{40}{\kilo\metre}$, and height $\SI{3}{\kilo\metre}$. The number of gridpoints in $x$ and $k$ is 800, and in $z$ is 257. The truncation limit $M$ (see \eqref{expansions}) is 200. Sensitivity tests were performed to ensure that increasing these resolutions does not impact the results.

The horizontal Prandtl number $Pr_h = \mathcal{A}_h/\mathcal{D}_h$ is assumed to be equal to one throughout - \citet{Shakespeare2017a} discuss the effect of non-zero Prandtl number on lee waves. Hereafter, we refer only to the viscosity $\mathcal{A}_h$, with the understanding that the diffusivity $\mathcal{D}_h$ varies similarly. 

\section{Results}\label{sec:results}
Results from the numerical solvers are now presented. First, the hydrostatic and non-rotating solution is shown to demonstrate the resonance and modification of generation that occurs when the horizontal group velocity is zero, as described in \S \ref{sec:groupvel} and \S \ref{sec:resonance}. Next, nonhydrostatic effects and rotation are introduced to the solution, and the results compared to the previous case. Finally, the effects of non-uniform stratification and velocity are shown.

\subsection{Hydrostatic and non-rotating solutions} \label{sec:results1}

\begin{figure} 
\centering
\includegraphics[width=5.3in]{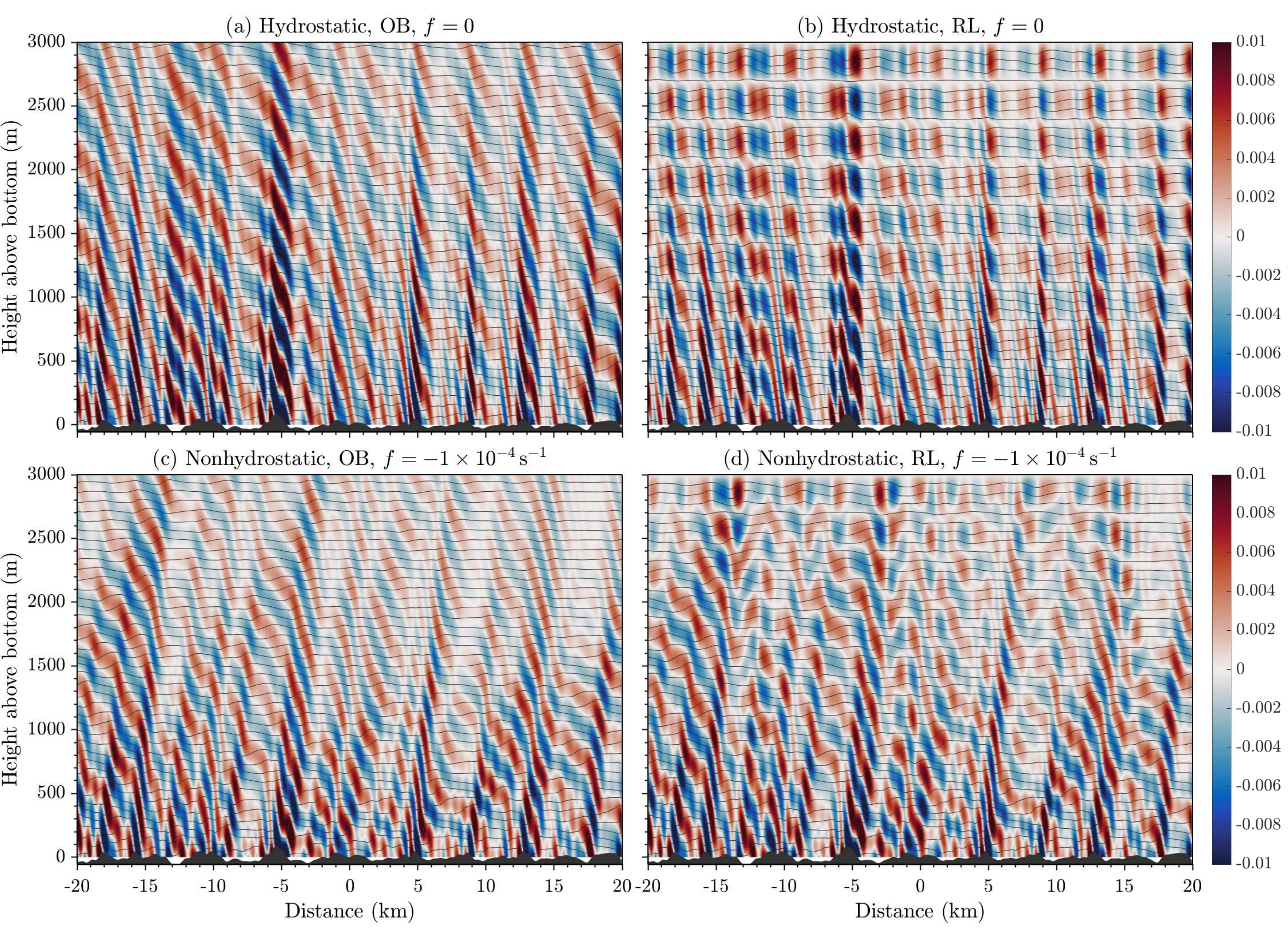} \\
\caption{Vertical velocity ($\SI{}{\metre\per\second}$) and isopycnals from the linear solver (a) with an open boundary (OB), $f =0$, hydrostatic, (b) with a rigid lid (RL), $f =0$, hydrostatic, (c) with an open boundary (OB), $f =  \SI{-1e-4}{\per\second}$, nonhydrostatic, (d) with a rigid lid (RL), $f =  \SI{-1e-4}{\per\second}$, nonhydrostatic. $\mathcal{A}_h =  \SI{1}{\metre\squared\per\second}$, $U = \SI{0.1}{\metre\per\second}$, $N = \SI{1e-3}{\per\second}$ for all cases. Topography $h(x)$ is shown, although it is applied in the linear approximation at its mean value of $z = 0$.} \label{fig5} 
\end{figure}

Figures \ref{fig5}a and \ref{fig5}b show the numerical linear solution for the vertical velocity field under the hydrostatic and non-rotating approximations with viscosity $\mathcal{A}_h = \SI{1}{\metre\squared\per\second}$. In figure \ref{fig5}a, there is an open boundary (OB) and waves can freely propagate out of the domain, whereas in figure \ref{fig5}b the rigid lid (RL) boundary condition is implemented. The reflection of waves and superposition back onto the wave field is clear, as is the well defined vertical wavenumber $m \sim N/U = \SI{0.01}{\per\metre}$, giving a vertical wavelength of $2\pi/m \sim \SI{628}{\metre}$. As discussed in \S \ref{groupvel}, when $f = \alpha = 0$, the horizontal component of group velocity is zero, as can be seen in the vertically radiating waves in figures \ref{fig5}a and \ref{fig5}b. As a result, waves reflected at the surface superimpose directly back onto the original wave field. Notice that near topography the solutions in figures \ref{fig5}a and \ref{fig5}b are similar since energy has been lost in the reflected wave, thus the solution consists mostly of the original upwards propagating component. 
\begin{figure} 
\centering
\includegraphics[width=5.3in]{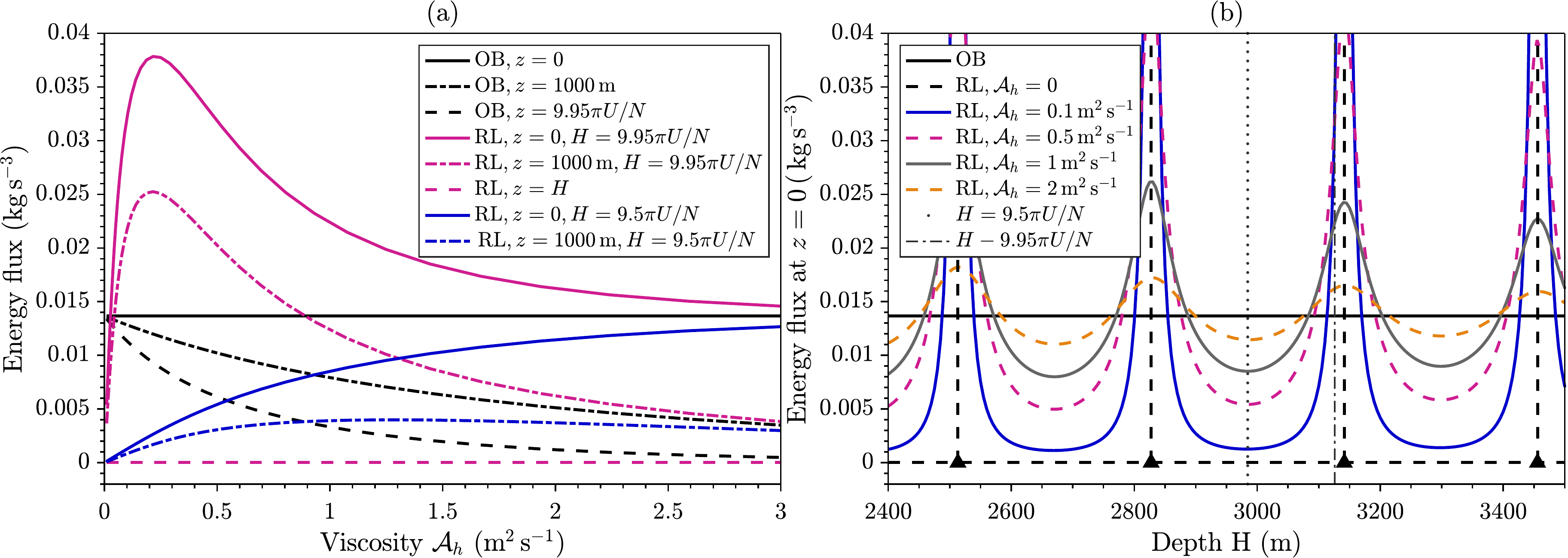} \\
\caption{(a) Horizontally averaged vertical energy flux at various heights for the open boundary (OB) and rigid lid (RL) hydrostatic and non-rotating solvers, against horizontal viscosity $\mathcal{A}_h$. (b) Horizontally averaged vertical energy flux at $z=0$ (proportional to wave drag) for several values of viscosity $\mathcal{A}_h$ against ocean depth $H$. Vertical dashed lines and triangles indicate the singularities $N^2H^2/U^2 = n^2\pi^2$. Other vertical lines indicate the values of $H$ shown in (a).$U = \SI{0.1}{\metre\per\second}$, $N = \SI{1e-3}{\per\second}$, $f = 0$ in both.} \label{fig6} 
\end{figure}

Section \ref{sec:resonance} describes how resonance can occur when $ N^2H^2/U^2= n^2\pi^2$, $n \in \mathbb{N}$, in the inviscid, non-rotating, hydrostatic scenario. Figure \ref{fig6} demonstrates this phenomenon with the given topography spectrum for varying viscosity $\mathcal{A}_h$. Figure \ref{fig6}a shows the lee wave energy flux at $z = 0, \SI{1000}{\metre}$, and  $H$ for the OB and RL solutions with $H = 9.95\pi U/N \simeq \SI{3126}{\metre}$ (constructive interference) and $H = 9.5\pi U/N \simeq \SI{2985}{\metre}$ (destructive interference). For the OB solution the energy flux at $z=0$ is almost independent of viscosity - it is modified slightly by the local viscous term at $z=0$ (not shown hereafter), but not the viscosity elsewhere in the domain since energy can only radiate away from the topography. As viscosity increases, the energy flux at $z = \SI{1000}{\metre}$ and the surface decreases as more energy is lost during propagation. At $\mathcal{A}_h = \SI{1}{\metre\squared\per\second}$, approximately $40\%$ of the wave energy dissipates in the bottom $\SI{1000}{\metre}$. 

For the RL solver, the results are markedly different for the two different domain heights $H$. In both cases, the energy flux is zero at $z=H$ due to the boundary condition, and zero when $\mathcal{A}_h = 0$, since there can be no steady state energy flux into lee waves without mixing and dissipation (see \eqref{energyint}). Equivalently, all upwards propagating energy flux is cancelled out by the reflected downwards component. However, for the near resonant case, when a small value of viscosity $\mathcal{A}_h = \SI{0.25}{\metre\squared\per\second}$ is introduced the energy flux at the topography increases to over 2.5 times that with no upper boundary. Now that there is no longer exact cancellation of the energy flux, constructive interference of the wave field initially allows the energy flux to increase with increasing $\mathcal{A}_h$. As viscosity increases further, energy loss of the reflected wave reduces the constructive interference and the energy flux at $z=0$ decreases, approaching that of the unbounded solution until it no longer `knows about' the boundary. The energy flux at $z=\SI{1000}{\metre}$ follows a similar pattern, approaching the OB flux as $\mathcal{A}_h$ increases. 

In contrast, the energy flux in the RL solution with $H =  9.5\pi U/N$ remains smaller than that in the OB solution throughout, since the effect of the boundary is to produce destructive interference with the original wave field. When the constructively interfering solution energy flux at $z=0$ peaks at $\mathcal{A}_h = \SI{0.25}{\metre\squared\per\second}$, the destructively interfering solution energy flux is $\sim 13$ times smaller. 

Figure \ref{fig6}b demonstrates this constructive/destructive behaviour of the wave field as $H$ varies. Again, the OB energy flux at $z=0$ is almost constant with changes in $\mathcal{A}_h$ and constant with changes in $H$. The RL energy flux at $z=0$ for $\mathcal{A}_h = 0$ is shown in black  dashes, with the triangles and asymptotes indicating the singularities at $ N^2H^2/U^2= n^2\pi^2$, $n = 8,9,10,11$. When $\mathcal{A}_h \neq 0$, the solutions become continuous with peaks at the singularities (constructive interference) and troughs halfway between (destructive interference). As $\mathcal{A}_h$ increases, the energy flux approaches the constant value of the OB solution. The values of $H$ in figure \ref{fig6}a are shown as vertical lines in figure \ref{fig6}b.
\begin{figure} 
\centering
\includegraphics[width=5.3in]{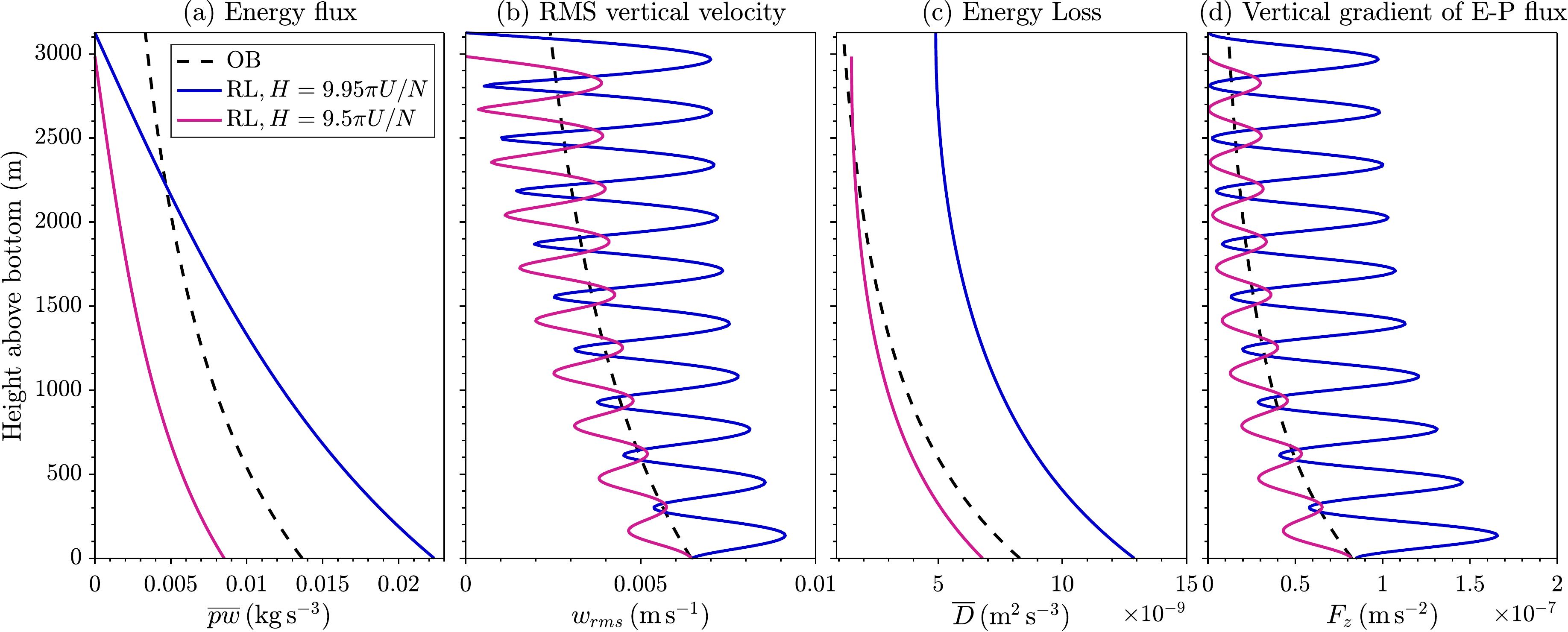} \\
\caption{ Horizontally averaged (a) energy flux (b) RMS vertical velocity, (c) energy loss, (d) vertical gradient of the E-P flux as a function of $z$ for the open boundary (OB) solver and the rigid lid (RL) solver with constructive and destructive interference. $\mathcal{A}_h = \SI{1}{\metre\squared\per\second}$, $f = 0$, $U = \SI{0.1}{\metre\per\second}$, $N = \SI{1e-3}{\per\second}$.} \label{fig7} 
\end{figure}

The vertical structure of the energy flux, RMS vertical velocity, energy loss, and vertical gradient of the E-P flux is shown in figure \ref{fig7} for the same cases as described in figure \ref{fig6} at $\mathcal{A}_h = \SI{1}{\metre\squared\per\second}$. Figure \ref{fig7}a again demonstrates the difference in energy fluxes with the boundary condition and height of domain. Figure \ref{fig7}b shows a periodic vertical structure in the RMS vertical velocity $w_{rms}$ of the RL solutions that does not exist in the OB solution,  due to the superposition of upwards and downwards propagating waves. This has the effect of enhancing the maximum $w_{rms}$ over a vertical wavelength, and decreasing the minimum, so that even in the destructive interference case where the energy flux in the RL solution is significantly smaller than the OB solution, the peak $w_{rms}$ is larger than that of the OB solution. 

The energy loss $\overline{D}$ (the sum of mixing and dissipation rate) is shown in figure \ref{fig7}c. In the RL case the vertical phases of the waves are aligned due to the surface boundary condition, and the mixing and dissipation rate individually have a sinusoidal structure out of phase with each other (not shown). This is due to the energy distribution in the wave alternating between kinetic and potential over a vertical wavelength. There is nearly 2.5 times more energy loss (normalised for domain height) when $H = 9.95\pi U/N$ (constructive interference) compared to when $H = 9.5\pi U/N$ (destructive interference). Comparing the RL and OB solutions for $H = 9.5\pi U/N$, it can be seen that energy loss in the RL solution is enhanced near the surface, suggesting that the upper boundary moves the distribution of wave energy (and therefore energy loss) higher up in the water column. 

The vertical gradient of the E-P flux is shown in figure \ref{fig7}d, representing the force on the flow due to wave breaking. In the RL cases, $F_z$ has a periodic structure in the vertical due to the wave interference, which would impact the feedback of the waves on the mean flow. The vertical integral of $F_z$ gives the total wave drag on the flow, hence the constructive interference produces a high drag state and the destructive interference a low drag state.

\subsection{Nonhydrostatic and rotating solutions} \label{sec:results2}
When rotation ($f \neq 0$) and nonhydrostatic ($\alpha = 1$ in \eqref{mom3}) effects are introduced, the resonance and interference effects described in \S \ref{sec:results1} are no longer as straightforward. As shown in \S \ref{sec:groupvel}, the horizontal component of the group velocity is now positive, allowing wave energy to travel downstream. Figures \ref{fig5}c and \ref{fig5}d show the vertical velocity field for the same background flow conditions and topography as figures \ref{fig5}a and \ref{fig5}b, but now with $f = \SI{-1e-4}{\per\second}$ and $\alpha = 1$. Waves now propagate downstream as well as vertically, and the resulting structure in the RL solution (figure \ref{fig5}d) is not as simple. However, the characteristic vertical phase lines and modal structure of the disturbances just below the surface caused by superposition of the reflected waves remain. 

Rotation reduces the generation of larger horizontal scale waves, and the dominant components of the wave field are therefore more easily dissipated than in the non-rotating solution shown in figures \ref{fig5}a and \ref{fig5}b. The vertical group velocity is also reduced by rotation (figure \ref{fig4}a), so the waves radiate more slowly away from the topography and lose more energy before reaching the surface. The wave field in the lower part of the domain of figure \ref{fig5}d therefore resembles the OB solution in figure \ref{fig5}c more closely than in the non-rotating solution, since the dominant wavelengths have lost more energy by the time they return to the topography.

\begin{figure} 
\includegraphics[width=5.3in]{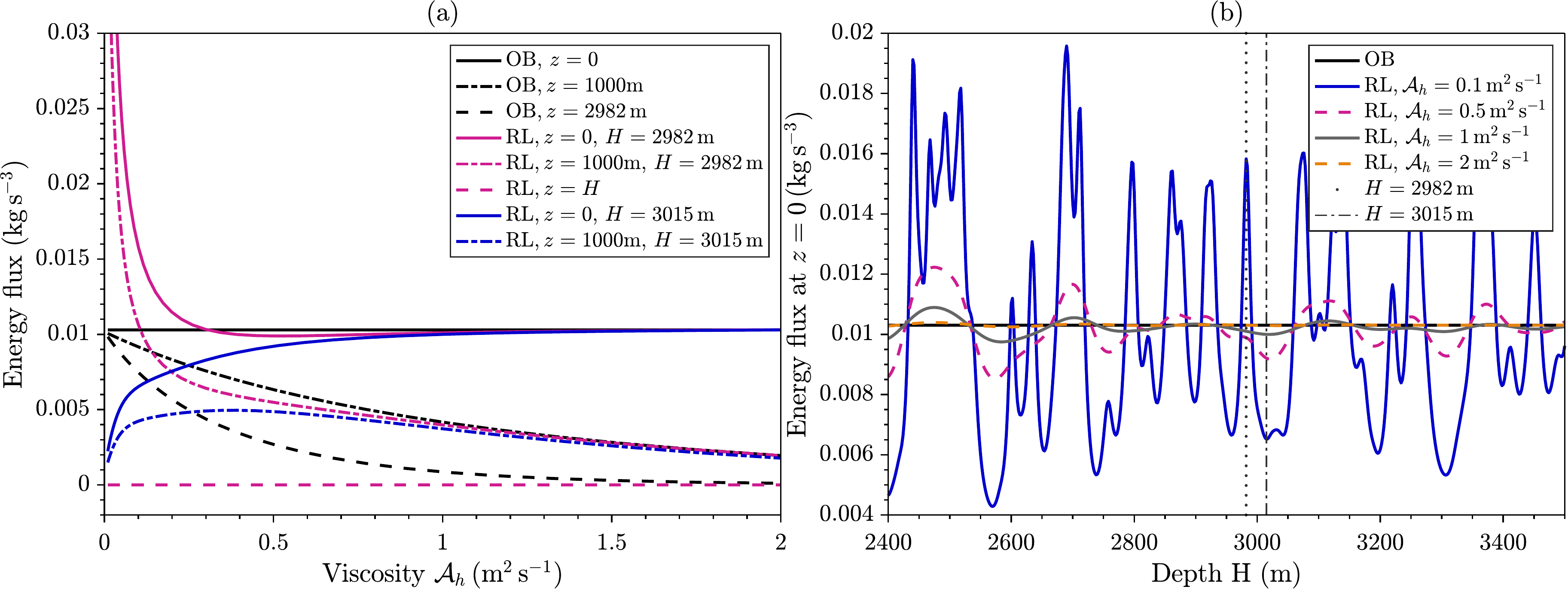} \\
\caption{(a) Horizontally averaged vertical energy flux at various heights for the open boundary (OB) and rigid lid (RL) nonhydrostatic and rotational solvers, against horizontal viscosity $\mathcal{A}_h$. (b) Horizontally averaged vertical energy flux at $z=0$ (proportional to wave drag) for several values of viscosity $\mathcal{A}_h$ against ocean depth $H$. Vertical lines indicate the values of $H$ shown in (a). $U = \SI{0.1}{\metre\per\second}$, $f = \SI{-1e-4}{\per\second}$, $N = \SI{1e-3}{\per\second}$ in both.}\label{fig8}
\end{figure}

Figure \ref{fig8} shows the same data as figure \ref{fig6}, now with rotation and nonhydrostaticity included, for two domain heights $H$ that have been picked to represent constructive and destructive interference of the new system. It is clear from figure \ref{fig8}b that the simple hydrostatic resonance has been replaced by multiple resonances where $|m(k)H| \simeq n\pi$ (c.f. \eqref{monochrom_sol}) for some $n \in \mathbb{N}$ and some $k$ in the spectrum $\hat{h}(k)$. As $H$ varies, the energy flux at the topography varies eratically as different wavenumbers $k$ in the topographic spectrum interfere constructively and destructively, with energy flux tending to that of the OB solution as $\mathcal{A}_h$ increases. The example values of $H$ in figure \ref{fig8}a are shown as vertical lines in figure \ref{fig8}b. At $H = \SI{2982}{\metre}$ there is net destructive interference, and energy fluxes are below those of the OB solution, whereas at $H = \SI{3015}{\metre}$ there is net constructive interference, and the energy flux is higher than the OB solution. The RL solutions tend to the OB solutions with increasing $\mathcal{A}_h$ more quickly than in figure \ref{fig6}, since the dominant wavelengths are shorter and decay faster. 

Importantly, since the horizontal group velocities are now positive so that the reflected wave does not directly superimpose onto the upwards propagating wave, the main reason for the modification of the bottom energy flux (and wave drag) with a reflecting upper boundary is the periodic nature of the topography used. The overlap parameter (defined in  \eqref{overlap}) for this set of parameters is above 2 for all wavenumbers (orange line in figure \ref{fig3}b). If the topography were isolated, the bottom energy flux may not be changed at all, dependent on the relevant overlap parameter. Even when there are near-resonances caused by constructive interference (peaks in figure \ref{fig8}b), they are smaller in amplitude than those in the hydrostatic, non-rotating case (figure \ref{fig6}) since the waves are dispersive, and the resonances occur for individual wavenumbers rather than the whole wave field. 

\begin{figure} 
\centering
\includegraphics[width=5.3in]{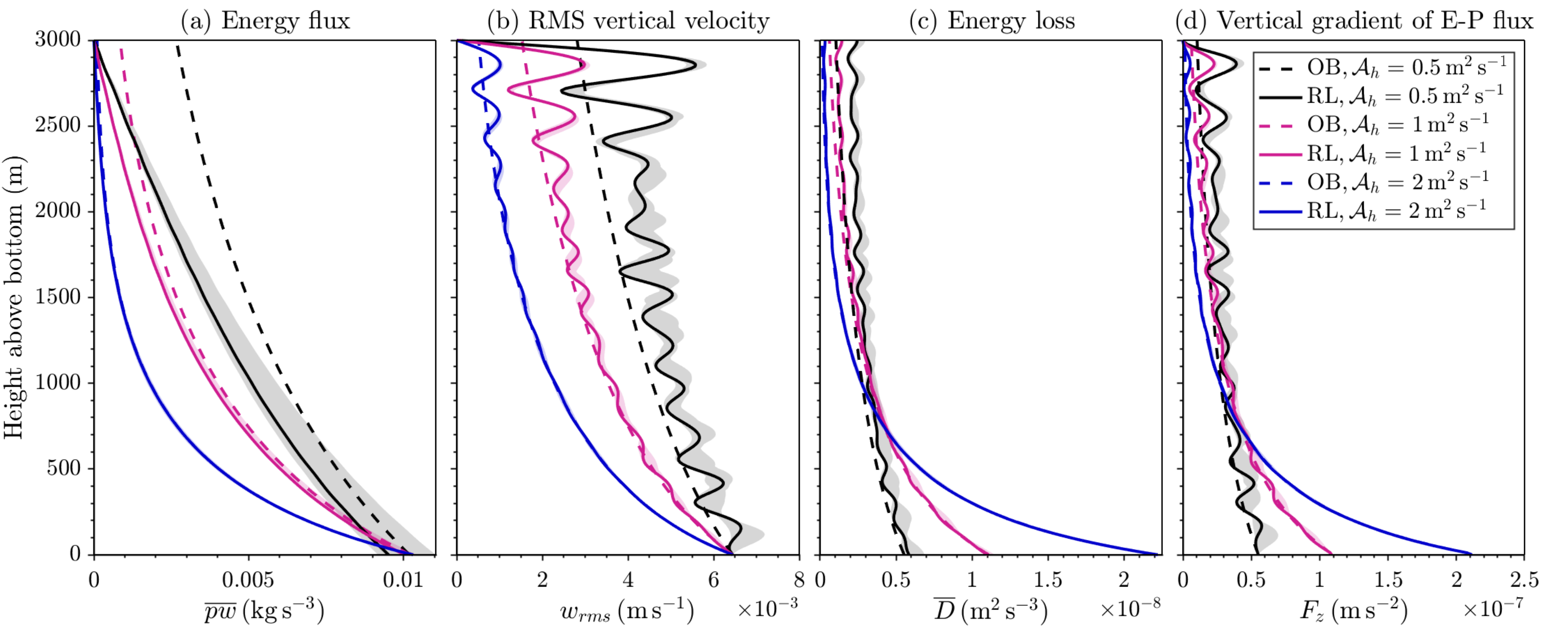} \\
\caption{Horizontally averaged (a) energy flux (b) RMS vertical velocity (c) energy loss (d) vertical gradient of E-P flux as a function of $z$ for the open boundary (OB) solver and the rigid lid (RL) solver for various values of $\mathcal{A}_h$. Results from the RL solver are shown as a range (shaded) of solutions with $H = 2900$ to $\SI{3100}{\metre}$ (with axes scaled onto $z\in [0,\SI{3000}{\metre}]$) to show the effect of constructive/destructive interference, and in solid at for $H = \SI{3000}{\metre}$. $f = \SI{-1e-4}{\per\second}$, $U = \SI{0.1}{\metre\per\second}$, $N = \SI{1e-3}{\per\second}$.}  \label{fig9} 
\end{figure}

From figure \ref{fig8}b, it is clear that even with periodic topography, when $\mathcal{A}_h  \gtrsim \SI{0.5}{\metre\squared\per\second}$, the constructive and destructive interferences with varying $H$ do not greatly affect the bottom energy flux; for $H > \SI{3000}{\metre}$ the change from the open boundary case is less than $10\%$ . $H$ is hereafter set to $\SI{3000}{\metre}$, although shaded regions in figures \ref{fig9}, \ref{fig11}, \ref{fig12}, \ref{fig13} and \ref{fig14} show the range of solutions for $H$ between $\SI{2900}{\metre}$ and $\SI{3100}{\metre}$ (with axes scaled onto $[0,\SI{3000}{\metre}]$), to indicate the extent of the interference. Figure \ref{fig9} shows the vertical structure of the fields as in figure \ref{fig7} for the nonhydrostatic and rotating RL and OB solutions at various values of $\mathcal{A}_h$.  The profiles of energy flux in figure \ref{fig9}a show, as expected, that the RL solution approaches the OB solution as $\mathcal{A}_h$ increases. They will be identical when the energy flux at $z=H$ in the OB solution is zero. 

As was found in the hydrostatic and non-rotating case in figure \ref{fig7}b, RMS vertical velocity profiles shown in figure \ref{fig9}b are generally enhanced for the RL compared to the the OB solution. $w_{rms}$ oscillates in $z$ due to the constructive and destructive interference of the wave field, with the maxima significantly larger than the OB solution, and the minima often larger too, especially in the lower viscosity cases. In particular, the subsurface maxima (located approximately $\pi U/2N \simeq \SI{157}{\metre}$ below the surface) are significantly larger than the OB solution at that level, $1.8-1.9$ times larger for each of the values of $\mathcal{A}_h$ here. They are also larger than the next deeper maximum below for each $\mathcal{A}_h$, and even larger than the RMS vertical velocities down to $z = \SI{700}{\metre}$  for $\mathcal{A}_h = \SI{0.5}{\metre\squared\per\second}$. The effect of the boundary is clearly to enhance the RMS vertical velocity in the upper ocean.

The energy loss shown in figure \ref{fig9}c is larger in the RL case than in the OB case for each value of $\mathcal{A}_h$. This is expected, since energy leaves the domain in the OB case, but must stay in the domain and be dissipated in the RL case. There is a subtlety in that in the RL case the bottom energy flux itself can be modified (see shading, and figure \ref{fig8}b and discussion), however the effect is not significant here. Consistent with the results of \citet{Zheng2019}, we find that the total energy loss over the water column is increased from the OB case, though the difference is not large, between $1\%$ for $\mathcal{A}_h = \SI{2}{\metre\squared\per\second}$ and $26\%$ for $\mathcal{A}_h = \SI{0.5}{\metre\squared\per\second}$. Assuming that the energy flux at the topography is unchanged by reflections, since $U$ is uniform with height the total energy loss for each case in figure \ref{fig9}c must be the same - but the OB solutions must be integrated to an infinite height to get this result. The main result of note is the difference in the distribution of energy loss in the water column when a RL is introduced - it is skewed towards the surface, with an increase of $45\%$ for $\mathcal{A}_h = \SI{2}{\metre\squared\per\second}$ and $70\%$ for $\mathcal{A}_h = \SI{0.5}{\metre\squared\per\second}$ in the top $\SI{1000}{\metre}$ compared to the OB case. 

The gradient of the E-P flux (figure \ref{fig9}d) has a similar structure to the energy loss (figure \ref{fig9}c). This is because $U$ is constant with height, and neglecting the effect of the upper boundary, both wave energy flux (the gradient of which for $U$ constant is given by $\overline{D}$ from \eqref{energyeqn}) and the E-P flux (with gradient $F_z$) decrease only due to mixing and dissipation. Equivalently to noting as above that the total height integrated energy loss should be the same, the total wave drag on the flow, given by the integral of $F_z$, should also be the same for the cases shown, though when restricting to $z \in [0,\SI{3000}{\metre}]$ the total wave drag in the RL solutions is larger than that of the OB solutions. 

For idealised, nonlinear, 2D, open boundary simulations with a similar topographic spectrum and flow parameters, \citet{Nikurashin2010b} found that $\sim 10\%$ of lee wave energy dissipated in the bottom $\SI{1}{\kilo\metre}$ for $Fr_L = 0.2$, and $\sim 50\%$ for $Fr_L \geq 0.5$ (representative of the Drake Passage). From figure \ref{fig8}a, these regimes would equate to $\mathcal{A}_h = \SI{0.2}{\metre\squared\per\second}$ and $\SI{0.7}{\metre\squared\per\second}$ respectively here. Of course, the change in implied turbulent viscosity between the two regimes is largely down to the nonlinearity and subsequent breaking for higher Froude number. This suggests that if a linear solution with a constant turbulent viscosity is to have any success in practise, it must be adjusted for the actual nonlinearity of the waves.

Comparing the energy loss (figure \ref{fig9}c) with the common parametrisation for the exponential vertical decay of lee wave energy dissipation \citep{Nikurashin2013,Melet2014}, the decay scale for the OB solver (calculated as the height above bottom at which energy loss is equal to $e^{-1}$ times its original value) is approximately $\SI{1700}{\metre}$ for $\mathcal{A}_h = \SI{0.5}{\metre\squared\per\second}$, $\SI{800}{\metre}$ for $\mathcal{A}_h = \SI{1}{\metre\squared\per\second}$, and  $\SI{400}{\metre}$ for $\mathcal{A}_h = \SI{2}{\metre\squared\per\second}$. Proposed values of the lee wave decay scale \citep{Nikurashin2013} range between $\SI{300}{\metre}$ and $\SI{1000}{\metre}$. This together with the comparison of the energy flux to the nonlinear simulations of \cite{Nikurashin2010b} suggests that we are in the correct parameter space for $\mathcal{A}_h$, and therefore that the upper boundary could have an influence on the wave field. We hereafter take $\mathcal{A}_h = \SI{1}{\metre\squared\per\second}$. 

\subsection{Non-uniform velocity and stratification}\label{sec:results3}

In reality, the assumption that background flow is uniform with height is unlikely to be valid when considering lee wave propagation throughout the entire water column. We now consider the impact of varying $U(z)$ and $N(z)$ on the lee wave field.

Unlike in the unbounded case, when solving the lee wave problem with a rigid lid upper boundary it is straightforward to solve with arbitrary mean flow velocity and stratification. Some constraints do apply, and we only consider velocity profiles $U(z) > 0$ such that $fU''(z) = 0$, so that the base state is effectively 2D (see  \S \ref{sec:basestate}), and $U'(z) > 0$, to avoid difficulties with critical levels (see \S  \ref{sec:critical}). Typical oceanic conditions are characterised by lower velocities at depth and larger velocities at the surface, so this scenario is realistic, although lee wave generation at locations of intensified bottom velocities may also be important \citep{Kunze2019}. The only constraint on the stratification $N^2$ is that $N^2 \geq 0$ so that the mean flow is statically stable, and viscosity is kept at $\mathcal{A}_h = \SI{1}{\metre\squared\per\second}$.  

\begin{figure}
\centering
\includegraphics[width=5.3in]{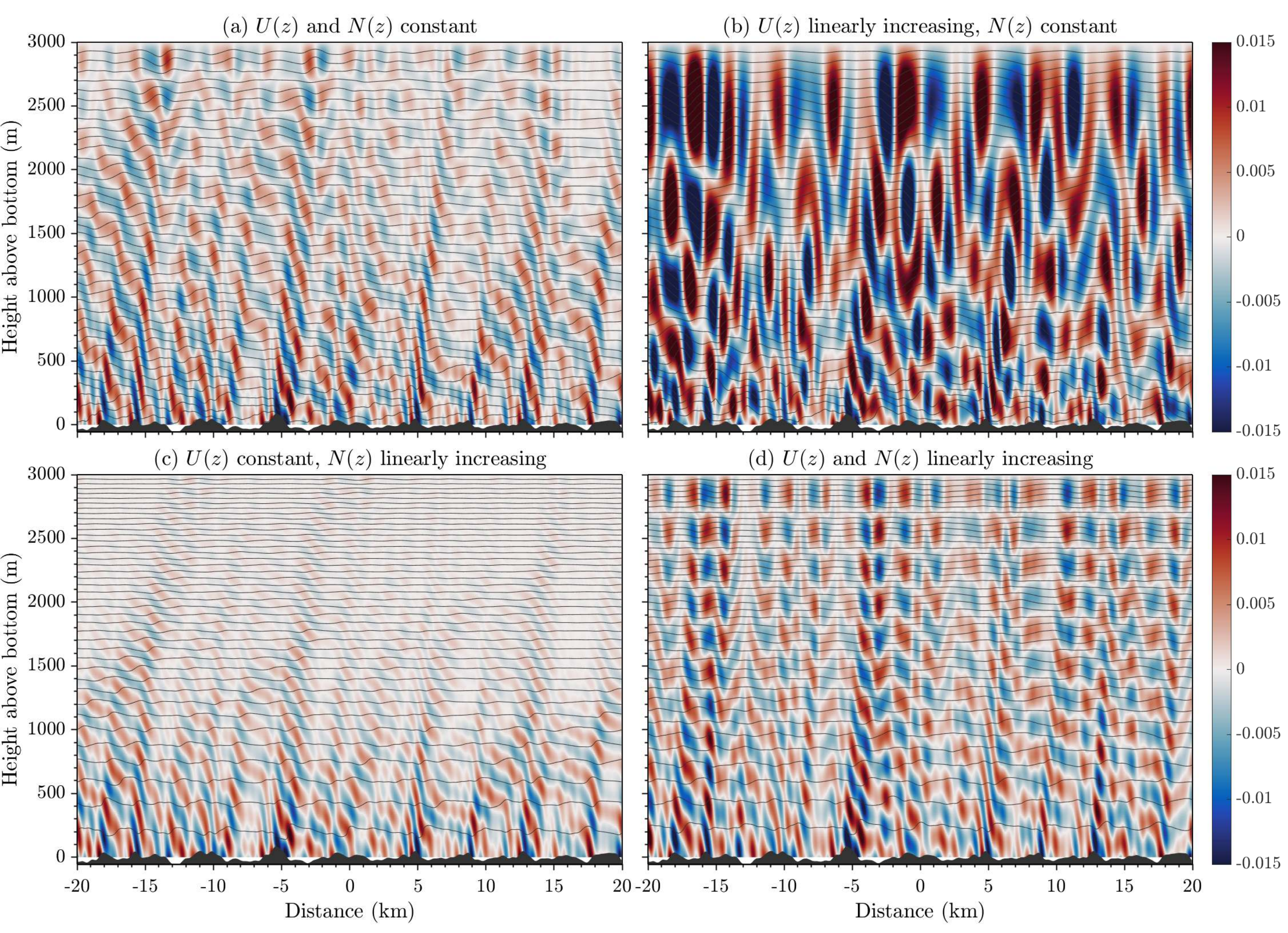} \\
\caption{Vertical velocity ($\SI{}{\metre\per\second}$) and isopycnals in the RL solver, with linear $U(z)$ and $N(z)$ with bottom values $U(0) = \SI{0.1}{\metre\per\second}$, $N(0) = \SI{1e-3}{\per\second}$, and (a) $U(H) = \SI{0.1}{\metre\per\second}$, $N(H) = \SI{1e-3}{\per\second}$ (b) $U(H) = \SI{0.3}{\metre\per\second}$, $N(H) = \SI{1e-3}{\per\second}$  (c) $U(H) = \SI{0.1}{\metre\per\second}$, $N(H) = \SI{3e-3}{\per\second}$  (d) $U(H) = \SI{0.1}{\metre\per\second}$, $N(H) = \SI{3e-3}{\per\second}$. $\mathcal{A}_h = \SI{1}{\metre\squared\per\second}$, $\alpha = 1$, and $f = \SI{-1e-4}{\per\second}$ in all cases. Topography $h(x)$ is shown, although it is applied in the linear approximation at it's mean value of $z = 0$.}  \label{fig10} 
\end{figure}

First, $U$ is varied linearly from $U(0) = \SI{0.1}{\metre\per\second}$ to $U(H) = 0.1, 0.2$ or $\SI{0.3}{\metre\per\second}$ with $N = \SI{1e-3}{\per\second}$ and $\mathcal{A}_h = \SI{1}{\metre\squared\per\second}$. The vertical velocity fields when $U(H) = 0.1$ and $\SI{0.3}{\metre\per\second}$ are shown in figures \ref{fig10}a and \ref{fig10}b respectively. It is clear that increasing $U(H)$ has a large effect on the wave field, with vertical velocities increased throughout the domain and increased dominant vertical and horizontal wavelengths as $U(z)$ increases. 

\begin{figure} 
\centering
\includegraphics[width=5.3in]{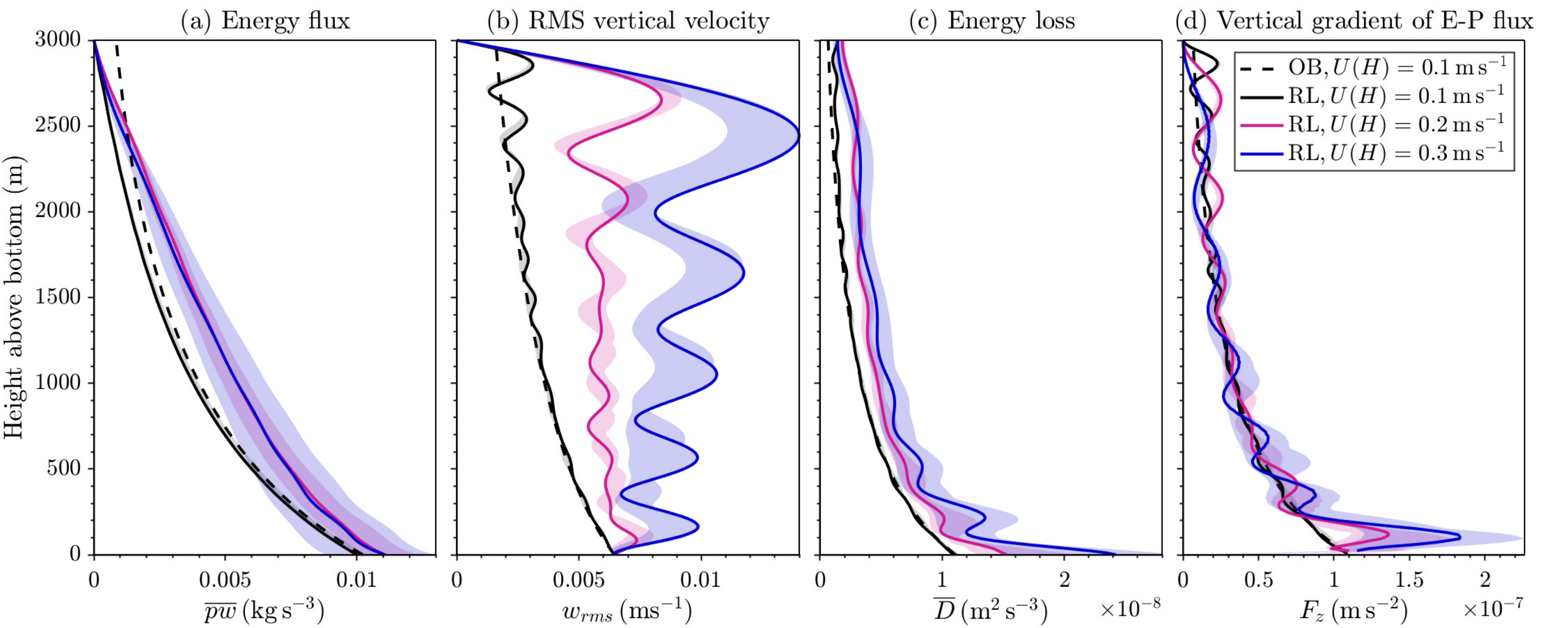} \\
\caption{Horizontally averaged (a) energy flux (b) RMS vertical velocity (c) energy loss (d) vertical gradient of E-P flux as a function of $z$ for the open boundary (OB) solver and the rigid lid (RL) solver for various $U(H)$, where $U(z)$ is linear and $U(0) = \SI{0.1}{\metre\per\second}$. Results from the RL solver are shown as a range (shaded) of solutions with $H = 2900$ to $\SI{3100}{\metre}$ (with axes scaled onto $z\in [0,\SI{3000}{\metre}]$) to show the effect of constructive/destructive interference, and in solid at for $H = \SI{3000}{\metre}$. $\mathcal{A}_h = \SI{1}{\metre\squared\per\second}$, $f = \SI{-1e-4}{\per\second}$, $N = \SI{1e-3}{\per\second}$.}  \label{fig11} 
\end{figure}

As before, waves are generated at the topography in the range $|f| < |U(0)k| < |N|$. This range can be visualised in figure \ref{fig3}b as the range of wavelengths for which the overlap parameter $\gamma$ is finite for $U = \SI{0.1}{\metre\per\second}$ (orange line). These are the only wavelengths that exist in the solution. However, as $U$ increases with height to $U(H) = \SI{0.3}{\metre\per\second}$, the range of wavenumbers that can propagate shifts to $|f| < |U(H)k| < |N|$, shown in figure \ref{fig3}b as the range for which $\gamma$ is finite for $U = \SI{0.3}{\metre\per\second}$ (pink line). Thus, wavenumbers $k$ such that $|N|/|U(H)| < |k| < |N|/|U(0)|$ must reach their turning levels and reflect downwards before reaching the surface. Therefore, the waves that reach the surface have larger horizontal wavelength, as seen in figure \ref{fig10}b, and decay more slowly as a result. The vertical group velocity of these waves also increases with increasing $U$, as shown (for constant $U$) in figure \ref{fig4}b, again reducing their energy loss whilst propagating through the water column. The largest wavelengths in the range also have a smaller overlap parameter and propagate at an angle closer to the vertical as $U$ increases (see figure \ref{fig3}b). As a result, they are more likely to interfere directly with their own upwards propagating components. These effects all result in increased interference, shown in figure \ref{fig11} as the shaded range becoming wider with increasing $U(H)$. 

The energy flux (figure \ref{fig11}a) at all levels is greater when $U(H) = 0.2$ and $\SI{0.3}{\metre\per\second}$ than when $U$ is uniform, except for near the topography when there is significant destructive interference. Aside from the ranges of the solutions (shaded), there is not a large difference between the energy fluxes in the cases $U(H) = 0.2$ and $\SI{0.3}{\metre\per\second}$, likely due to the effect of certain larger wavenumbers reaching their turning levels at $|U(z)k| = |N|$ and reflecting before reaching the surface, decreasing upwards energy flux at higher levels. From \eqref{EPenergy}, the energy flux $\overline{pw}$ is expected to increase with increasing $U(z)$ when there is no energy loss and the E-P flux is conserved. However, the RL solution constrains the energy flux to vanish at the surface, thus the convexity of the energy flux in $z$ will be determined by the balance between the gradient of the E-P flux due to energy loss and reflection, and the gradient of $U(z)$.

The most obvious result of increasing the velocity with height is the increase in the RMS vertical velocity, shown in figure \ref{fig11}b. Despite the large ranges due to interference, increasing $U(H)$ clearly increases $w_{rms}$ over the whole water column. The subsurface maximum when $U(H) = \SI{0.3}{\metre\per\second}$ is 4.5 times as large as that when $U(H) = \SI{0.1}{\metre\per\second}$, and over twice as large as its bottom value. The vertical wavelength also clearly increases with increasing $U$.

As discussed by \citet{Kunze2019}, lee waves can exchange energy with the mean flow due to conservation of wave action $E/kU$, where $E$ is the energy density \citep{BrethertonGarrett1969}. For a given wavenumber $k$, when there is no energy lost to dissipation, the wave energy density will therefore increase with height when $U$ increases with height. Here, the energy lost to dissipation means that the wave action is not conserved, but we still expect the wave energy to increase in the upper water column when $U(z)$ increases with height compared to when $U(z)$ is constant. Energy loss would also be expected to increase along with wave energy density for a given wavenumber. Figure \ref{fig10}c demonstrates that energy loss over the water column does generally increase with increasing $U(H)$. However, because energy at some wavenumbers no longer reaches the surface, having reached the corresponding turning level at $|U(z)k| = |N|$, the increase in wave energy (and energy loss) in the upper ocean is not as great for the multichromatic spectrum of waves as for a single component that reaches the surface. The waves that do reach the surface also experience less energy loss due to their larger horizontal scale.
 
Although the energy loss is generally greater when $U$ increases with $z$, figure \ref{fig11}d shows that the vertical gradient of the E-P flux is not, aside from the changes due to interference. Since $F$ is conserved when there is no mixing or dissipation, the flux does not increase due to interaction with the shear. Neglecting wave interference from surface reflections at the topography, the total wave drag $\rho_0F(0)$ depends only on the local fields, and thus remains constant with changes in $U$ with height. However, the total mixing and dissipation need not, since the waves can gain energy from the mean flow during propagation. 

\begin{figure} 
\centering
\includegraphics[width=5.3in]{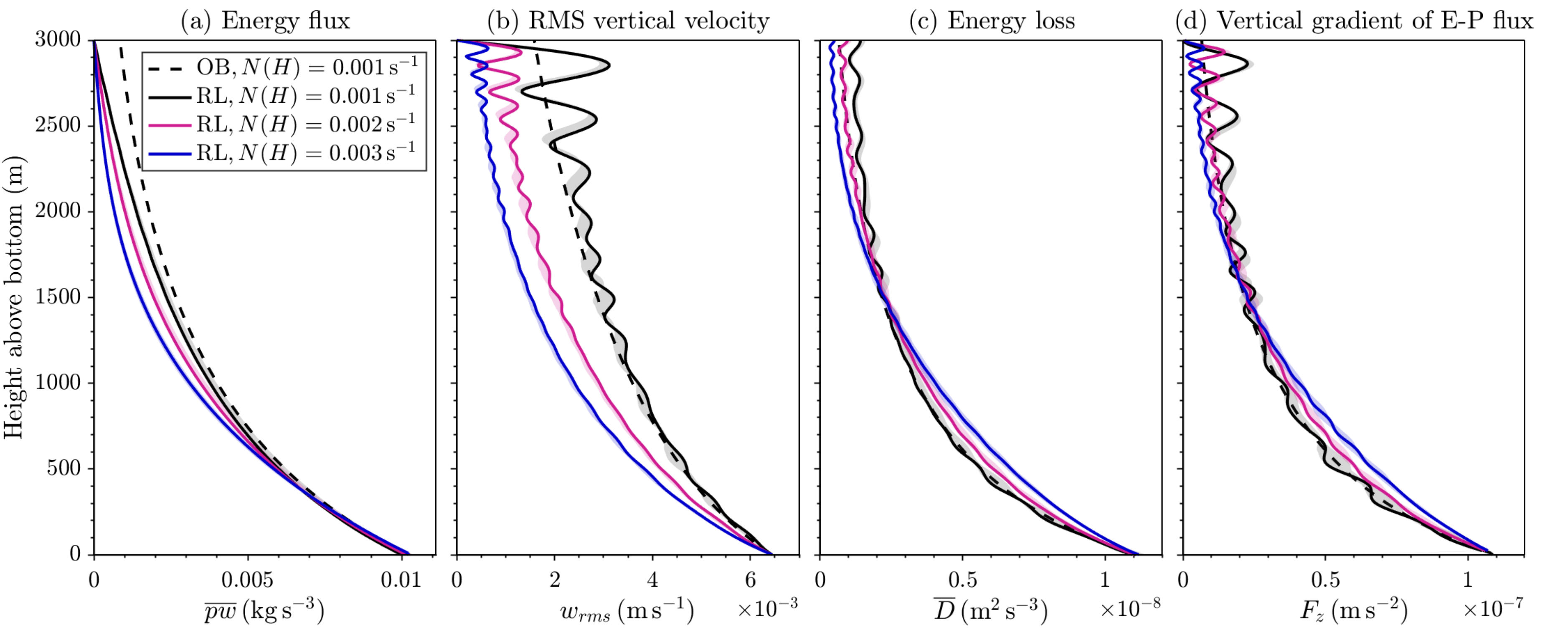} \\
\caption{Horizontally averaged (a) energy flux (b) RMS vertical velocity (c) energy loss (d) vertical gradient of E-P flux as a function of $z$ for the open boundary (OB) solver and the rigid lid (RL) solver for various $N(H)$, where $N(z)$ is linear and $N = \SI{1e-3}{\per\second}$. Results from the RL solver are shown as a range (shaded) of solutions with $H = 2900$ to $\SI{3100}{\metre}$ (with axes scaled onto $z\in [0,\SI{3000}{\metre}]$) to show the effect of constructive/destructive interference, and in solid at for $H = \SI{3000}{\metre}$. $\mathcal{A}_h = \SI{1}{\metre\squared\per\second}$, $f = \SI{-1e-4}{\per\second}$, $U = \SI{0.1}{\metre\per\second}$.}  \label{fig12} 
\end{figure}

We now consider the effect of increasing the stratification $N(z)$ on the lee waves in the RL solver. The vertical velocity field is shown in figure \ref{fig10}c. The vertical wavelengths are clearly reduced as $N$ decreases, since the vertical wavenumber $m \sim N/U$. The vertical velocities are also reduced higher in the water column when compared to figure \ref{fig10}a. Figure \ref{fig12} shows the vertical profiles as in \ref{fig11}. It is immediately clear from the lack of shaded area that in the cases shown, constructive/ destructive interference does not greatly affect the amplitude of the solutions, and to a decreasing extent for increasing $N(H)$. This is because the vertical group velocity \eqref{groupvel} scales as $1/N$, so the waves lose more energy during their propagation and thus interact less. If vertical viscosity and diffusivity were implemented, the smaller vertical wavelengths associated with increased $N$ would dissipate even more quickly. 

Figure \ref{fig12}b shows clearly that the effect of increasing $N$ with height is to reduce the vertical velocities. Since varying $N$ with height does not affect the energy flux in the same way as changing the velocity does (c.f. \eqref{EPenergy}), the other results in figure \ref{fig12} are easily interpreted. The increase in energy loss associated with reduced group velocity when $N$ is increasing causes a reduction in energy flux (figure \ref{fig12}a), and a skewing of energy loss (figure \ref{fig12}c) and gradient of the E-P flux (figure \ref{fig12}d) towards the lower part of the domain. Note from \eqref{energyint} that the vertically integrated energy loss is constant with changing $N(H)$ (when $U(z)$ is constant), as is the total wave drag force on the flow, given by the integral of $F_z$ (figure \ref{fig12}d). 

Next, we present results for simultaneously varying $U(z)$ and $N(z)$, keeping their ratio constant at $U(z) = 100\times N(z)$, so that the vertical scales of the lee wave field are comparable. This is a fairly realistic scenario for the ocean, where both $U$ and $N$ can be expected to increase with height above bottom. Figure \ref{fig10}d shows the vertical velocity field when both $U$ and $N$ triple between the bottom and the surface. The vertical wavelengths are comparable with figure \ref{fig10}a as expected, however, the vertical velocities are intensified, and there is more interference of the upwards and downwards propagating waves.

\begin{figure} 
\centering
\includegraphics[width=5.3in]{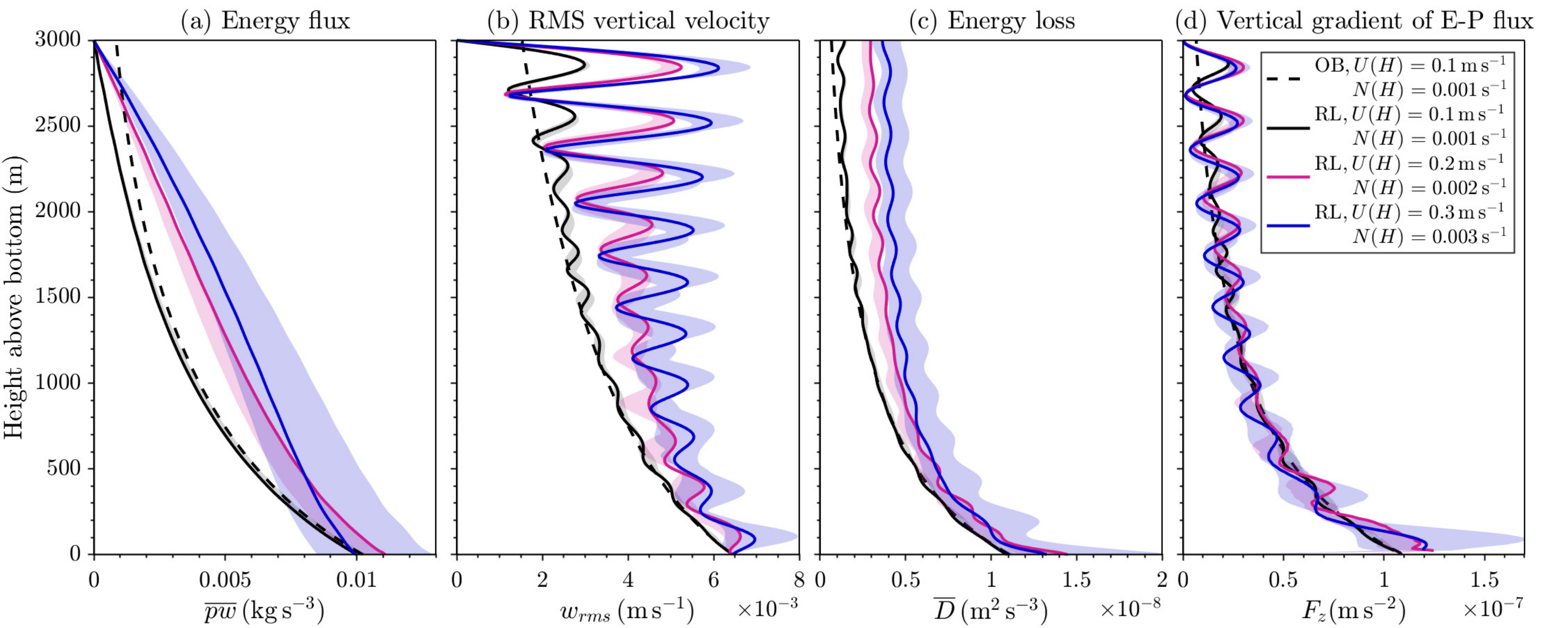} \\
\caption{Horizontally averaged (a) energy flux (b) RMS vertical velocity (c) energy loss (d) vertical gradient of E-P flux as a function of $z$ for the open boundary (OB) solver and the rigid lid (RL) solver for various $U(H)$, where $U(H)$ is linear and $U(H) = 100N(H)$. Results from the RL solver are shown as a range (shaded) of solutions with $H = 2900$ to $\SI{3100}{\metre}$ (with axes scaled onto $z \in [0,\SI{3000}{\metre}]$) to show the effect of constructive/destructive interference, and in solid at for $H = \SI{3000}{\metre}$. $\mathcal{A}_h = \SI{1}{\metre\squared\per\second}$, $f = \SI{-1e-4}{\per\second}$.}  \label{fig13} 
\end{figure}

The energy flux with height is shown in figure \ref{fig13}a. As in figure \ref{fig11}, the ranges associated with interference are larger as $U(H)$ increases. In general, upper ocean energy flux increases with increasing $U(H)$ and $N(H)$. The gradient of the E-P flux (figure \ref{fig13}d), has a similar structure in the vertical for each case, with a greater range due to interference for larger values of $U(H)$ as expected. The distribution of the forcing on the mean flow is therefore largely unchanged by increasing $U$ and $N$ with height. 

\begin{figure}
\centering
\includegraphics[width=5.3in]{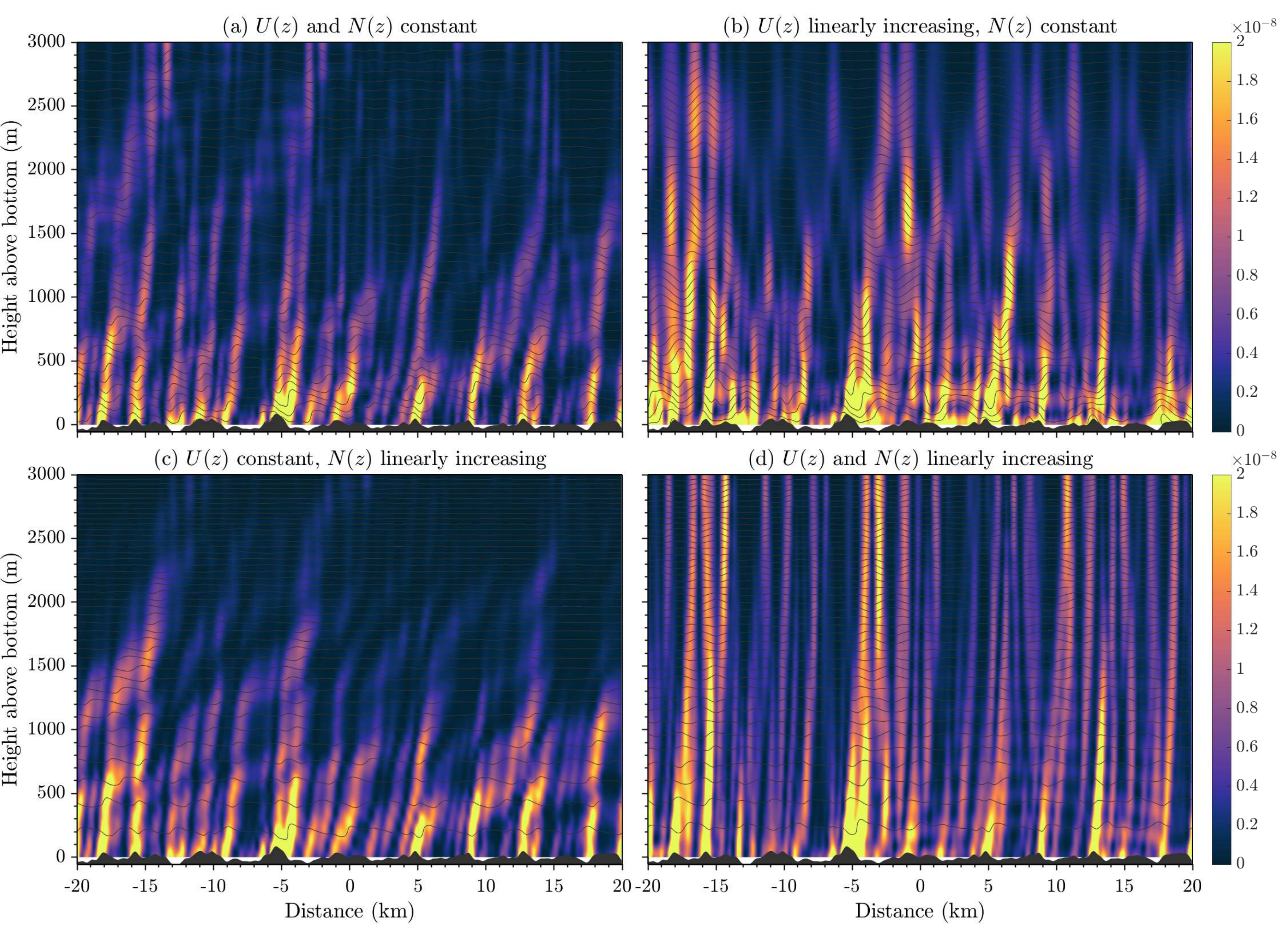} \\
\caption{Energy loss $D$ ($\SI{}{\metre\squared\per\second\cubed}$) and isopycnals in the RL solver, with the same mean flow and parameters described in figure \ref{fig10}. } \label{fig15} 
\end{figure}

The upper ocean energy loss (figure \ref{fig13}c) increases with increasing $U(H)$ and $N(H)$, which can be explained as before by conservation of wave action as $U$ increases, transferring energy to the wave field and increasing the wave energy density and hence energy loss. However, unlike the $U$ increasing case (figure \ref{fig11}c), the energy loss at most heights is now strictly increasing with $U(H)$, since the constant ratio of $U/N$ means that the range of radiating wavenumbers does not change with height, thus no wavenumber reaches a level of internal reflection. The result of this is that the energy loss in the upper ocean is significantly enhanced when $U$ and $N$ increase with height. The energy loss in the upper $\SI{1000}{\metre}$ is three times larger when $U$ and $N$ approximately triple with height than when they are uniform throughout the water column (both with a RL). The change in energy loss with height for the various background flows is also illustrated in figure \ref{fig15}. As we have seen, energy loss increases slightly with height with respect to the uniform fields when $U$ increases with height, and decreases when $N$ increases with height. The combination of increasing both $U$ and $N$, however, allows the waves to stay in their radiating range and gives the maximum upper ocean energy loss. 

Another interesting result is the large increase in RMS vertical velocity with height when $U$ and $N$ increase together (figure \ref{fig13}b), suggesting that the increase of $w_{rms}$ due to increasing $U$ is dominant over the decrease in $w_{rms}$ due to increasing $N$ (see figures \ref{fig11}b and \ref{fig12}b). The subsurface maximum of $w_{rms}$ when $U(H) = \SI{0.3}{\metre\per\second}$ is twice as large as that when $U(H) = \SI{0.1}{\metre\per\second}$, and nearly 4 times as large as $w_{rms}$ in the OB solution at the same height. The impact of the boundary is substantial, with the variation in $w_{rms}$ over a vertical wavelength due to superposition increasing with increasing $U(H)$ and $N(H)$.

\begin{figure} 
\centering
\includegraphics[width=5.3in]{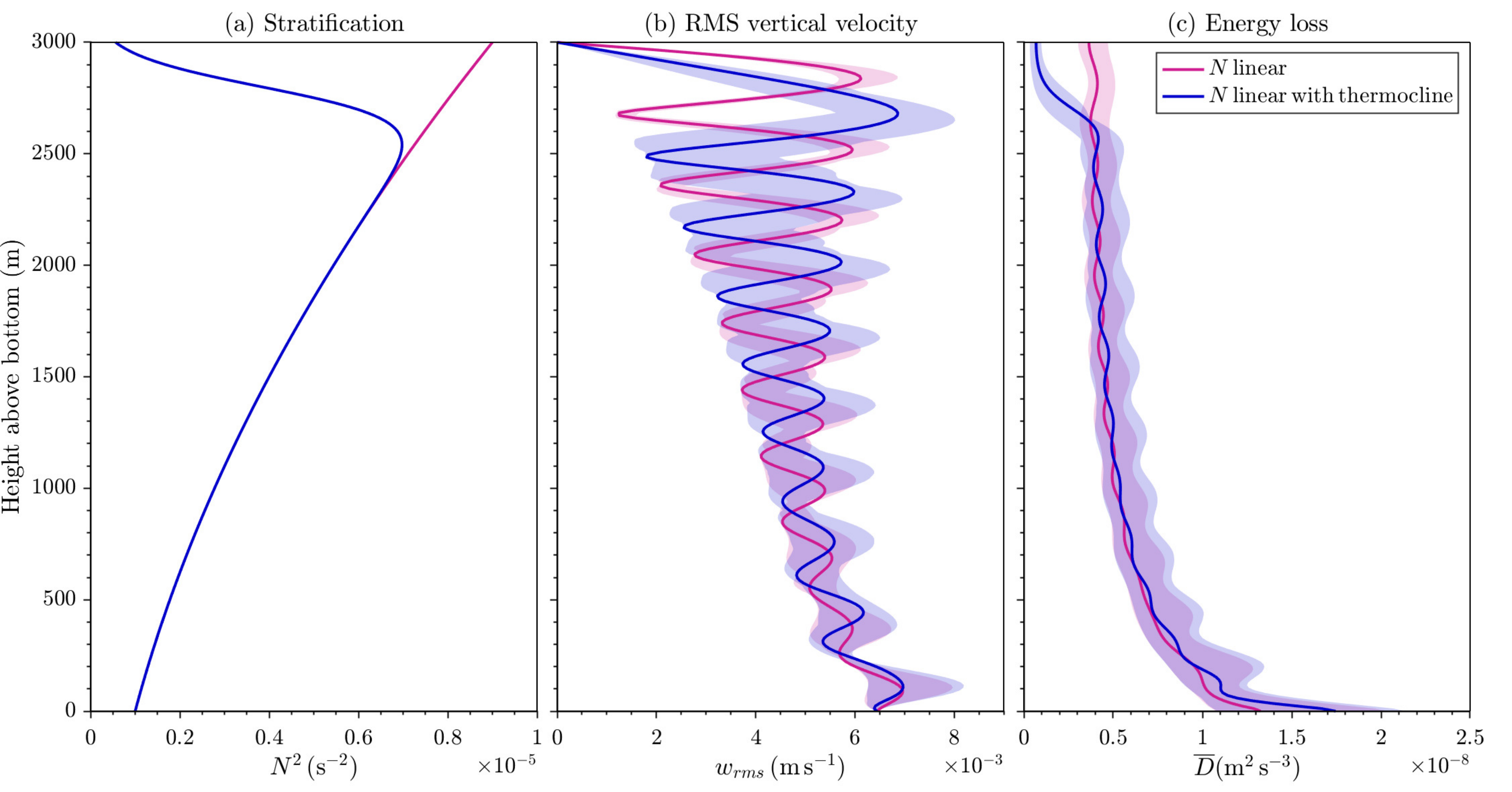} \\
\caption{(a) Stratification, (b) RMS vertical velocity, and (c) energy loss  as a function of $z$ for the RL solver. $U$ is linear with $U(z) = 0.1(1 + 2z/H)$. The linear $N$ profile is $N(z) = 0.001(1 + 2z/H)$, and with the thermocline included is $N(z) = 0.001(1 + 2z/H)\sqrt{(1+\tanh(18-20z/H))/(1+\tanh(18))}$. Results from the RL solver are shown as a range (shaded) of solutions with $H = 2900$ to $\SI{3100}{\metre}$ (with axes scaled onto $z \in [0,\SI{3000}{\metre}]$) to show the effect of constructive/destructive interference, and in solid at for $H = \SI{3000}{\metre}$. $\mathcal{A}_h = \SI{1}{\metre\squared\per\second}$, $f = \SI{-1e-4}{\per\second}$.}  \label{fig14} 
\end{figure}

Finally, we consider the effect of a more realistic stratification in the upper ocean. Typically, there exists a maximum of stratification at the thermocline, and a mixed layer at the surface where stratification is near zero. We use a simplified example stratification representative of the mean stratification in realistic Drake Passage simulations (which are themselves constrained by observed hydrographic information), having a maximum at around $\SI{500}{\metre}$ depth and decreasing to zero at the surface \citep{Mashayek2017a}. In reality, the stratification can have a second sharp maximum below the thin surface mixed layer dependent on seasonality, but the deeper thermocline is a persistent feature. For comparison with the previous experiments, $N$ is linear (and $N^2$ quadratic) at depth, and modified using a $\tanh$ function to create the thermocline. Figure \ref{fig14}a shows the profiles of $N^2$ used, and $U$ is linearly increasing from $U(0) = 0.1$ to $U(H) =  \SI{0.3}{\metre\per\second}$. The effect of the drop in stratification at the surface, as might be expected from figure \ref{fig12}b, is to further enhance the subsurface peak in RMS vertical velocity (figure \ref{fig14}b). Although $w_{rms}$ increases, the buoyancy and horizontal velocity perturbations decrease with $N^2$ near the surface (not shown), leading to a decrease in total flow energy and energy loss (figure \ref{fig14}c). The combined effect of increasing velocity with height above bottom, the reflecting upper boundary, and a near surface decrease in stratification all act to increase the subsurface peak in RMS vertical velocity. 

\section{Conclusions}\label{sec:discussion}

Lee waves generated by stratified geostrophic flow over topography play an important role in the buoyancy and momentum budgets of the ocean, causing diapycnal mixing and drag on the mean flow when they break. Occurring at the sub-gridscale of global models, they require parametrisation to represent their effect on the mean flow.

Linear theory with constant background velocity and stratification and a radiating upper boundary has often been used to predict the generation rate of lee waves. However, although this approximation may be sufficient locally to estimate the generation of lee waves, it does not allow any deductions on their propagation throughout the water column and eventual dissipation or re-absorption to the mean flow. The mean velocity and stratification in typical oceanic flows varies by up to an order of magnitude between the abyssal ocean and the surface, and the ocean surface is poorly represented by a radiating boundary condition, instead acting to reflect incident lee waves.

Motivated by high resolution realistic simulations of the Drake Passage, a region of high lee wave generation, we developed a theory for lee waves with an air-sea boundary, variable background velocity and stratification, and a representation of energy lost to dissipation and mixing. The structures observed in the simulations agree qualitatively with our theoretical predictions, and reconciling the two will be the subject of a follow up study. 

We find that allowing lee waves to reflect at the surface has the potential to substantially modify the lee wave field, increasing vertical velocities and mixing and dissipation, especially in the upper ocean where shear and stratification are typically enhanced. 

Allowing waves to reflect at the surface allows interference between the upwards and downwards propagating components, and this can modify the lee wave generation itself. Under certain conditions, this may manifest as a resonance of the system - although rotation and non-hydrostaticity act to lessen this effect.

The upper boundary alone acts to enhance near surface vertical velocities, and shift the energy loss of the lee wave field higher in the water column.  However, the impact of our full water column view of lee waves is most significant when combined with non-uniform background flows, as are typical of realistic ocean conditions.

When the background velocity increases with height above the sea floor, as is often the case in wind driven geostrophic flows, we find that the impact of the reflection from the surface increases and that the lee wave vertical velocities are significantly enhanced in the upper ocean. The lee wave drag is largely unchanged, but the energy in the lee wave field, and hence the energy lost to mixing and dissipation, increases since energy transfers from the sheared mean flow to the waves.  If the stratification also increases with height such that $U/N$ remains fairly constant, the waves that are generated at topography are all able to reach the surface, increasing the upper ocean wave energy and energy loss. The inclusion of a weakly stratified surface mixed layer acts to enhance near surface vertical velocities further, and reduces near surface energy loss. Therefore, parametrising the effect of lee waves propagating through changing background flows may be essential for correctly estimating their impact on mixing.

The simplifications made in this study leave some questions as to the applicability of these results to the real ocean. In particular, although linear lee wave approximations have been shown to give good agreement with nonlinear simulations under certain conditions, the wave interactions discussed here that cause modification to wave drag and energy flux could be significantly altered by nonlinear topography. The assumption of linearity also has consequences for the wave-mean flow interaction, particularly when the background flow changes with height and energy transfers between waves and mean flow. Here, the mean flow is forced to remain constant, whereas in practise the mean flow would lose energy to the lee waves.

The contribution of time dependent components of the background flow, including tides, could change the nature of the wave interactions. Internal tides could contribute to lee wave dissipation via wave-wave interactions, and the unsteady nature of tides themselves could impact the lee wave reflection and superposition through modification of the large scale flow. However, lee waves that can interact with the surface are most likely to occur when geostrophic flow speeds are high, thus at relevant locations the geostrophic flow is likely to dominate the tidal flow.

The constructive and destructive interference of the wave field may be overestimated due to the use of a periodic topography consisting of a finite sum of topographic components. It is likely that for a realistic topography, where the peaks of topography that generate lee waves are isolated and at different heights, this effect is substantially reduced. The effect of 3D topography could also alter the results, and this could be investigated in the linear framework by extending the solver.

We implemented a horizontal viscosity and diffusivity in place of the full Laplacian parametrisation of lee wave energy loss for mathematical simplicity, which becomes unrealistic when the vertical scale of lee waves changes substantially over the depth of the water column. Breaking due to instabilities of the lee waves themselves is not explicitly accounted for, since the viscosity and diffusivity are constant with height. The appropriate values of viscosity and diffusivity should also vary with the nonlinearity of the waves themselves, and this could be especially important when the background flow changes with height, potentially changing the stability of the waves. However, the resulting energy loss was shown to agree with previous nonlinear simulations with similar topography and background flow \citep{Nikurashin2010b}.

A rigid lid boundary condition has been used here, justified by the lack of impact of a free surface on the structure of the waves in the interior. However, predictions of the sea surface height imprint of these waves could be made within our theory. This could perhaps eventually allow observational diagnostics - the steady nature of lee waves would make them a good candidate for satellite observation. Modern satellite observations are fast approaching the $\mathcal{O}(\SI{1}{\kilo\metre})$ horizontal resolution and $\mathcal{O}(\SI{1}{\cm})$ precision that would be necessary to detect the very largest waves \citep{Neeck2012}. Satellite sun glitter images can also qualitatively be used to diagnose lee wave surface signatures \citep{deMarez2020}. The rigid lid condition would however be appropriate for modelling under-ice lee waves, whose surface normal stress could play a role in sea ice or ice shelf dynamics.

The results of this study indicate that the reflection of lee waves at the ocean surface and their presence in the upper ocean cannot always be neglected, especially when the mean flow is surface intensified. Climate model parametrisations may need to take into account the impact of changing background mean flows and surface reflections in order to correctly estimate the vertical structure of mixing and dissipation. Enhanced upper ocean mixing could have important consequences for tracer transport between the surface and interior ocean. The dynamics of the near surface wave field and its interaction with surface submesoscales should also be investigated further, since the horizontal lengthscales are very similar. Further studies will aim to verify the theory developed here against realistic nonlinear simulations, and investigate the impact of these waves on surface processes.

\section*{Code availability}
The code for the numerical solver used in this study will be made available upon publication. 

\section*{Acknowledgments}
L.B. was supported by the Centre for Doctoral Training in Mathematics of Planet Earth, UK EPSRC funded (grant no. EP/L016613/1), and A.M. acknowledges funding from the NERC IRF fellowship grant NE/P018319/1.\\

Declaration of Interests. The authors report no conflict of interest.

 \bibliographystyle{jfm}

\end{document}